\documentclass[12pt]{article}
\usepackage{amsbsy,amssymb,amsmath,bm,epsfig,rotating,wrapfig,appendix}
\usepackage{cancel}
\usepackage{graphicx}
\usepackage{multirow}
\usepackage{fancyhdr}

\numberwithin{equation}{section} \textwidth 155mm \textheight
225mm  
\begin{document}

\title{
\begin{flushright}
\small ITP-UU-08/61 \\
\small SPIN-08/48 \\
\small arXiv:0810.3329 \\
$\quad$ \\
$\quad$ \\
$\quad$ \\
\end{flushright}
Axion Stabilization in Type IIB Flux Compactifications}

\author{Kiril Hristov\footnote{k.p.hristov@uu.nl}\\
{\small\em Institute for Theoretical Physics and Spinoza
Institute,}\\
{\small\em Utrecht University, Leuvenlaan 4, 3584 CE Utrecht, the
Netherlands}}
\date{}

\maketitle

\begin{abstract}
A scenario for stabilization of axionic moduli fields in the
context of type IIB Calabi-Yau flux compactifications is discussed
in detail. We consider the case of a Calabi-Yau orientifold with
$h^{1,1}_- \neq 0$ which allows for the presence of $B_2$ and
$C_2$-moduli. In an attempt to generalize the KKLT and the Large
Volume Scenario, we show that these axions can also be stabilized
- some already at tree level, and others when we include
perturbative $\alpha'$-corrections to the K\"{a}hler potential $K$
and nonperturbative D$3$-instanton contributions to the
superpotential $W$. At last, we comment on the possible influence
of worldsheet instantons on the process of moduli stabilization.
\end{abstract}

\section{\label{intro}Introduction}

In the past few years there has been great research interest in
the field of string phenomenology, dealing with the question of
stabilizing moduli fields at desirably high masses (for a
comprehensive review see e.g. \cite{rev1,rev2}). This was
initiated by the KKLT scenario \cite{kklt} which suggested a way
to obtain stabilized vacua from type IIB string theory building on
earlier works such as \cite{earlier,giddings-kachru}. Presently,
one can find many extensions and improvements of the original
idea, the most notable and well established of which is the Large
Volume Scenario (LVS) \cite{lvs1,lvs2}. It builds up on the KKLT
solutions by allowing for non-supersymmetric vacua and by
including perturbative corrections to the tree-level K\"{a}hler
potential computed in \cite{BBHL}. Up to now the LVS has passed
many consistency checks (e.g. \cite{jumpingloops}), but there is
nevertheless much space for improvement. The stabilization of the
K\"{a}hler moduli requires manifolds with negative Euler number,
as well as non-perturbative effects which appear only if certain
conditions are satisfied \cite{witten,renata,flux_instantons}. And
it is of course desirable to have a working recipe also for the
other cases. The process of uplifting to a Minkowski or de Sitter
vacuum also needs to be understood better because at present it
seems that unnatural fine tuning of parameters is necessary.

In this paper we propose another extension, namely the
stabilization of moduli fields that arise from the two-form R-R
and NS-NS fields in type IIB string theory. These are usually
neglected in the literature, where the main focus is on
stabilizing the volume of the underlying manifold to large enough
values. Here we will argue that the stabilization of these so
called axionic or non-geometric moduli is an important step in
drawing the full picture. We show that these axions may lead to
changes in the process of stabilization of the manifold volume and
the other moduli. Additional motivation for considering them are
the possible cosmological consequences from their existence - they
are good candidates for driving inflation as recently suggested in
\cite{eva}. Here we will try to put these considerations on a firm
ground, first showing explicitly the existence of a large number
of flux compactifications in F-theory that include axions. These
are afterwards translated to the type IIB compactifications on
Calabi-Yau orientifolds, where the analysis of moduli
stabilization is better understood. Then we will be able to
generalize the existing stabilization techniques in order to
accommodate for the new moduli.

For this reason we first try to give a brief introduction to type
IIB flux compactifications in section \ref{flux}, including the
axions in the general discussion. In section \ref{tree_level} we
discuss the stabilization procedure at tree level. We then show
how stabilization changes after including perturbative and
D$3$-instanton corrections, in sections \ref{alpha'-corrections}
and \ref{D3-instantons} respectively. We comment on both the
supersymmetric (KKLT) and non-supersymmetric (LVS) type of vacua.
Based on \cite{mirror,uu,gopakumar} we are also able to estimate
the importance of the worldsheet instantons on the moduli
potential in section \ref{worldsheet instantons} and we see that
the $B_2$-moduli might substantially alter the moduli
stabilization procedure in the large volume limit. We conclude by
listing the possible applications of the axion moduli and
suggestions for further research in section \ref{CONC}. Some of
the more technical calculations used in the main text are carried
out in the appendices.

\section{Flux Compactifications in Type IIB String Theory}\label{flux}

We will first briefly review flux compactifications of type IIB
string theory establishing the basic conventions and equations
that will be used later.

The particle content of the type IIB supergravity is derived from
the massless spectrum of the corresponding superstring type. The
fermionic part consists of two left-handed Majorana-Weyl
gravitinos and two right-handed Majorana-Weyl dilatinos. As
supersymmetry holds and all fermionic degrees of freedom
correspond exactly to bosonic ones, specifying either part of the
effective action completely determines the other one. In this case
there are $32$ supersymmetry generators, i.e. we are in the case
of $\mathcal{N} = 2$ supergravity in $10$ dimensions. We will then
concentrate on the bosonic part from here on, keeping in mind the
fermionic counterparts. In the bosonic spectrum we have NS-NS and
R-R bosons. The NS-NS bosons are the metric $g_{M N}$, a two-form
$B_2$ (with corresponding field strength $H_3 = d B_2$) and the
dilaton $\phi$. The R-R sector consists of corresponding form
fields $C_0, C_2$, and $C_4$, the latter having a self-dual field
strength $F_5$ (also $F_1 = d C_0$ and $F_3 = d C_2$). In order to
obtain four dimensional models with $\mathcal{N} = 1$ we need to
compactify the theory on Calabi-Yau orientifold where fluxes are
turned on under the conditions:

\begin{equation}\label{fluxes}
  \frac{1}{(2 \pi)^2 \alpha'} \int_{\Sigma_{\alpha}} F_3 =
  n_{\alpha} \in \mathbb{Z}, \qquad \frac{1}{(2 \pi)^2 \alpha'} \int_{\Sigma_{\beta}} H_3 =
  m_{\beta} \in \mathbb{Z},
\end{equation}
with $\Sigma_{\alpha, \beta}$ three-cycles on the manifold. The
resulting metric becomes a warped product of flat four-dimensional
spacetime and conformally Calabi-Yau orientifold.

The compactification as described in this picture essentially
requires a Calabi-Yau three-fold with $O3/O7$ orientifold planes,
$D3/D7$ branes and the fluxes from (\ref{fluxes}). There is
however another description of the same physical situation if one
considers F-theory on an elliptically fibered Calabi-Yau four-fold
\cite{sen}. There one needs to add only $D3$ branes and fluxes and
the theory is equivalent to the one of type IIB flux
compactification. Since in this way one obtains the orientifold
"for free" without the need of explicitly constructing $O3/O7$
projection as in the type IIB picture, the F-theory approach is
widely used for realistic constructions. The rules of translating
between the two pictures are simple to use. A detailed summary can
be found in section 4.1 of \cite{denef}. Here we will need to know
that ($h^{1,1} (CY_4) - 1$) corresponds to $h^{1,1}_+$ and
$h^{2,1} (CY_4)$ to $h^{1,1}_-$, where $h^{1,1}_{+,-}$ are the
Hodge numbers on the Calabi-Yau orientifold counting the even
resp. odd parts of the ($1,1$)-homology under the orientifold
projection. The tadpole cancellation condition that needs to be
satisfied in the F-theory picture is:
\begin{equation}\label{tadpole}
    \frac{1}{(2 \pi)^2 \alpha'} \int H_3 \wedge F_3 + N_{D3} -
    N_{\bar{D}3}= \frac{\chi (CY_4)}{24},
\end{equation}
where $\chi$ is the Euler number of the four-fold. In the type IIB
picture this number effectively collects the contribution to the
$D3$ brane charge from the orientifold planes and the $D7$ branes.
Clearly, $\chi (CY_4)$ needs to be divisible by $24$, which puts a
restriction on the space of elliptic four-folds that can be used
for compactification (not too strict one since $\chi (CY_4) = 48 +
6 (h^{1,1} + h^{3,1} - h^{2,1})$).

The resulting effective field theory corresponds to a standard
$\mathcal{N} = 1$ supergravity with number of scalar (moduli)
fields counted by the Hodge numbers. The KKLT and LVS scenarios,
as well as the vast literature on the subject of type IIB moduli
stabilization, focus the attention on breaking the no-scale
structure of the potential and on stabilizing the K\"{a}hler
moduli at a value where the internal manifold has a large volume
as consistency requires. In this process the non-geometric
K\"{a}hler moduli are usually completely disregarded and assumed
non-existent. This is only justified in special cases for
orientifold projections where $h_{-}^{1,1} = 0$, as otherwise we
have additional moduli coming from the 2-form fields $B_2$ and
$C_2$ of the type IIB low energy effective action. One can find
many examples of Calabi-Yau four-folds leading to both $h^{1,1}_-
= 0$ and $h^{1,1}_- \neq 0$ (cf. \cite{internet} or Table B.4 of
\cite{elliptic} - keep in mind that $h^{1,1}_- = h^{2,1} (CY_4)$).

For a generic manifold ($h^{1,1}_- \neq 0$), the moduli to be
stabilized in the theory are the axio-dilaton $\tau = C_0 + i e^{-
\phi}$ (from here on referred to simply as dilaton), $h^{2,1}_-$
complex scalars $z^i$ parametrizing the size of the surviving
three-cycles appearing in (\ref{fluxes}), the K\"{a}hler moduli:
\begin{equation}\label{geometric moduli}
    J = v^{\alpha} (x) \omega_{\alpha} (y), \qquad \alpha = 1,...,h_{+}^{(1,1)},
\end{equation}
and the corresponding axionic moduli $\rho_{\alpha}$ from the
four-form $C_4$:
\begin{equation}\label{non-geometric 4-form moduli}
    C_4 = \rho_{\alpha} (x) \tilde{\omega}^{\alpha} (y), \qquad a = 1,...,h_{+}^{(1,1)},
\end{equation}
with $\{ \tilde{\omega}^{\alpha} \}$ the basis of harmonic
$(2,2)$-forms, dual to the $(1,1)$ basis $\{ \omega_{\alpha} \}$
that is even under the orientifold projection. The additional
moduli entering the effective four-dimensional field theory
because of $h^{1,1}_-$ are:
\begin{equation}\label{non-geometric moduli}
    B_2 = b^a (x) \omega_a (y), \qquad C_2 = c^a (x) \omega_a (y), \qquad a = 1,...,h_{-}^{(1,1)},
\end{equation}
where $\{ \omega_a \}$ is the basis of harmonic $(1,1)$ forms that
are odd under the orientifold projection. In the above formulae,
$x$ denotes the four-dimensional space-time where all the moduli
(and we) live, and $y$ are the coordinates on the compact
six-dimensional internal manifold.

With these definitions, the K\"{a}hler metric on the space of
moduli fields is given in terms of the reduced complex structure
coordinates coming from the explicit manifold and in terms of the
dilaton, the K\"{a}hler and the axionic moduli arranged as follows
\cite{soft,grimm-louis}:
\begin{eqnarray}\label{real Kaehler coordinates}
\nonumber & \tau = C_0 + i e^{- \phi}, \qquad \qquad G^a = c^a - \tau b^a, \\
T_{\alpha} & = \frac{3 i}{2} \rho_{\alpha} + \frac{3}{4}
\kappa_{\alpha} (v) + \frac{3 i}{4 (\tau - \bar{\tau})}
\kappa_{\alpha a b} G^a (G - \bar{G})^b,
\end{eqnarray}
where $\kappa_{\alpha} (v) \equiv \kappa_{\alpha \beta \gamma}
v^{\beta} v^{\gamma}$, i.e. it is just a four-cycle volume (with a
different normalization compared to the standard literature, used
for simplicity). In this notation,
\begin{equation}
\kappa \equiv \kappa_{\alpha} v^{\alpha} = 6 V_{CY} =
\kappa_{\alpha \beta \gamma} v^{\alpha} v^{\beta} v^{\gamma},
\end{equation}
where $V_{CY}$ is the volume of the manifold already after the
orientifold projection. The numbers $\kappa_{\alpha \beta \gamma}$
and $\kappa_{\alpha a b}$ are the usual Calabi-Yau intersection
numbers after performing the orientifold projection. As explained
in \cite{grimm-louis,grimm}, in the process of orientifolding
consistency requires that only the intersection numbers with even
number of Latin indices are non-zero. This means that for all
$\alpha, \beta, a, b, c$, $\kappa_{\alpha \beta a} = \kappa_{a b
c} = 0$ has to hold. The explicit construction of orientifolds
with such properties might not be straightforward. However, we
need not worry about this issue since orientifolding is performed
implicitly from the F-theory picture and thus consistency is
guaranteed.

The standard $\mathcal{N}=1$ F-term potential\footnote{Here we do
not add D-terms that are also allowed in $\mathcal{N}=1$
supergravity.  These generally appear whenever there are charged
chiral fields in the effective action. In principle this happens
when one tries to reproduce the MSSM by adding D$7$-branes
\cite{moster}, but here we strictly concentrate on moduli
stabilization and therefore neglect the possibility for a D-term
potential.} for the moduli fields is given by:
\begin{equation}\label{full_potential}
  V = e^K \left( K^{I \bar{J}} D_I W D_{\bar{J}} \bar{W} - 3 |W|^2
  \right),
\end{equation}
where the indices $I, J$ run over all chiral fields (the ones
defined through (\ref{real Kaehler coordinates}) together with the
complex structure moduli $z^i$), the matrix $K^{I \bar{J}}$ is the
inverse of the K\"{a}hler metric $K_{I \bar{J}} \equiv
\partial_{I}
\partial_{\bar{J}} K$, and $D_I W =
\partial_I W + \partial_I K \cdot W$. Here, the K\"{a}hler potential $K$ and the superpotential $W$ are
functions of the moduli fields in a particular way that will be
discussed separately in the following sections. Once $K$ and $W$
are known, the moduli potential $V$ can be calculated and the
minima to which the moduli fields roll down and get stabilized can
be found in principle.

From the above definitions, we see that:
\begin{equation}\label{diff_coords}
\kappa_{\alpha} = \frac{2}{3} (T_{\alpha} + \bar{T}_{\alpha}) -
\frac{i}{2 (\tau - \bar{\tau})} \kappa_{\alpha a b} (G-\bar{G})^a
(G-\bar{G})^b.
\end{equation}
Had we assumed that $h_{-}^{1,1} = 0$ the additional
$G^a$-dependent term would vanish and everything would be the same
as in \cite{kklt}, so we see that the results in the literature
are consistent with the neglect of the non-geometric moduli.
However, if we really want to stabilize all moduli in the generic
case where $h_{+}^{1,1} \sim h_{-}^{1,1} \sim \mathcal{O}(100)$ we
need to use the coordinate basis given by (\ref{real Kaehler
coordinates}). We will then describe in detail what happens in
this case and show how all these moduli will be eventually
stabilized in a manner similar to the KKLT and LVS procedures. In
what follows we separately discuss the resulting moduli potential
and its stabilization for the tree-level case, and for the cases
with added perturbative $\alpha'$-corrections to $K$ and then
D$3$-instantons to $W$. In the end we will be also able to draw
conclusions on how the addition of worldsheet instanton
corrections to the K\"{a}hler potential can influence the
stabilization process.

Note that once we derive the moduli potential from the K\"{a}hler
metric in the basis of chiral fields $\{ \tau, T_{\alpha}, G^a
\}$, we will be able to switch to the basis of real scalars
$\{C_0, \phi, v^{\alpha}, \rho_{\alpha}, b^a, c^a \}$ using
(\ref{real Kaehler coordinates}). It will turn out that
minimization of the potential is easier in this new basis since
the volume of the Calabi-Yau $V_{CY} = \frac{1}{6} \kappa_{\alpha
\beta \gamma} v^{\alpha} v^{\beta} v^{\gamma}$ will depend only on
the two-cycle moduli $v^{\alpha}$ and not on the other scalars. Of
course, once having stabilized all scalars one can always switch
back to the initial chiral fields where the metric on the moduli
space takes a simpler form. For additional clarity we present
Table \ref{table}, listing the chiral and real fields that appear
in this work, their multiplicity and associated indices.

\begin{table}
\begin{center}
\begin{tabular}{|c|c|c|c|} \hline
Index &  Chiral fields  &   Real scalars   & Values  \\
    \hline
- & $\tau$ & $C_0, \phi$ & 1 \\
$i$   & $z^i$ & $Re (z^i), Im (z^i)$ & $1,..., h^{2,1}_-$ \\
$\alpha$   & $T_{\alpha}$ & $v^{\alpha} \leftrightarrow \kappa_{\alpha}, \rho_{\alpha}$ & $1,..., h^{1,1}_+$\\
$a$   & $G^a$ & $b^a, c^a$ & $1,..., h^{1,1}_-$ \\
    \hline
\end{tabular}
\caption{Multiplicity of chiral and real moduli.}\label{table}
\end{center}
\end{table}

\section{\label{tree_level}Tree level} At tree level, in four dimensional $\mathcal{N} = 1$ supergravity,
the K\"{a}hler potential is (see e.g. \cite{giddings-kachru})
\begin{equation}\label{tree_kaehler}
  K = -\ln[i \int_{CY} \Omega (z) \wedge \bar{\Omega} (\bar{z})] -
  \ln (-i (\tau - \bar{\tau})) - 2 \ln(V_{CY}),
\end{equation}
where the $V_{CY} = \frac{\kappa_{\alpha} v^{\alpha}}{6}$ has to
be regarded as a function of the true K\"{a}hler coordinates
(\ref{real Kaehler coordinates}). For $\kappa_{\alpha}$ we use
(\ref{diff_coords}), while $v^{\alpha}$ can only be written in
terms of the chiral fields implictly by inverting the quadratic
relation $\kappa_{\alpha} = \kappa_{\alpha \beta \gamma} v^{\beta}
v^{\gamma}$. The superpotential at tree level is independent of
the K\"{a}hler and axionic moduli and is given by the famous
Gukov-Vafa-Witten \cite{GVW} flux superpotential
\begin{equation}\label{tree_W}
  W (z^i, \tau) = \int_{CY} \Omega (z^i) \wedge (F_3 - \tau H_3).
\end{equation}
A detailed calculation of the K\"{a}hler potential and the
superpotential was carried out in
\cite{giddings-kachru} and generalized to all orientifolds in \cite{grimm-louis}, both quantities follow from the $\mathcal{N} = 2$
dimensional reduction of the low-energy effective action before
orientifolding. The full moduli potential can be calculated from
Eq. (\ref{full_potential}). It is important here to stress that
the potential at tree-level is positive semi-definite. This is not
directly obvious from the expression, but is nevertheless true as
it comes from the reduction of the $\mathcal{N} = 2$, where it is
manifestly positive definite (c.f. App. A.2 of
\cite{giddings-kachru}). This means that any full minimum of the
potential will be at $V = 0$ and local minima (if any) could be
only of de Sitter type (at $V > 0$).

With this information, we can now try to investigate the explicit
form of the potential. The somewhat involved calculation of the
K\"{a}hler metric and its inverse are carried out in App.
\ref{tree appendix} and the results are in exact accordance with
those in \cite{grimm-louis,smet}. One of the main results is given
by the simple expression
\begin{equation}\label{no-scale potential - 4}
  K^{A \bar{B}} K_A K_{\bar{B}} = 4,
\end{equation}
where the indices $A,B$ run over $\tau, T_{\alpha}, G^a,$ and not
over the complex structure moduli $z^i$. One can roughly break
this sum into two contributions - a part in which the dilaton is
involved plus a part coming only from the $G^a$'s and the
$T_{\alpha}$'s as given by (\ref{break into sums}). This will be
helpful when we want to search for minima of the moduli potential.

\begin{figure}
  \begin{center}
  \includegraphics[width=8.0cm]{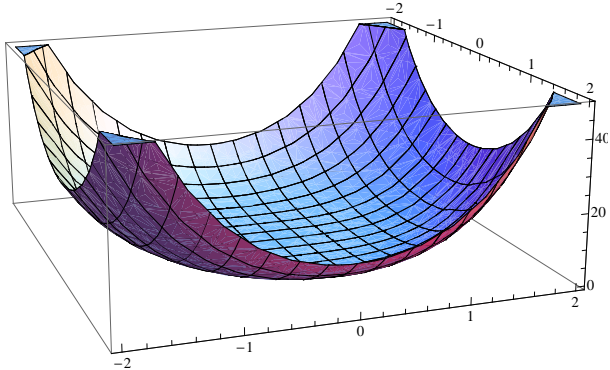}
  \end{center}
  \caption{The form of the tree-level potential in the $b^a$-directions  for $h^{1,1}_- = 2$.}\label{hat}
\end{figure}

The moduli potential (\ref{full_potential}) can now be calculated
easily from (\ref{inverse full kaehler metric}) and
(\ref{tree_W}), but its minima cannot be found analytically and
depend on the specific model. The only class of controlled minima
is realized when we stabilize the complex structure moduli and the
dilaton to a supersymmetric minimum, $D_{z^i} W = D_{\tau} W = 0$
- the same procedure used in the KKLT and LVS. Imposing $D_{z^i} W
= D_{\tau} W = 0$ results in stabilizing all $z^i$'s and $\tau$ to
be some function of $\frac{\kappa_{\alpha a b} v^{\alpha} b^a
b^b}{V_{CY}}$ as it appears in $K_{\tau}$
(\ref{partial_der_full_kaehler}). When $\kappa_{\alpha a b}
v^{\alpha} b^a b^b = 0$ this dependence vanishes and $z^i, \tau$
are stabilized to constants as in the original KKLT. The moduli
potential after fixing $D_{z^i} W = D_{\tau} W = 0$ becomes
\begin{equation}\label{tree potential with
v,b unstabilized} V = \frac{e^K e^{- 2 \phi} |W|^2}{4 V_{CY}^2}
(\kappa_{\alpha a b} v^{\alpha} b^a b^b)^2.
\end{equation}
It is manifestly positive semi-definite once more. Clearly we can
reach the global minimum $V = 0$ if $\kappa_{\alpha a b}
v^{\alpha} b^a b^b = 0$. In the initial Calabi-Yau three-fold,
$\kappa_{\hat{\alpha} \hat{\beta} \hat{\gamma}} v^{\hat{\alpha}}$
has a signature $(1,h^{1,1}-1)$ \cite{candelas} (here we use the
convention $(+,-)$ for matrix signature). After the projection,
$\kappa_{\alpha \beta \gamma} v^{\alpha}$ is with signature
$(1,h_{+}^{1,1}-1)$ and $\kappa_{\alpha a b} v^{\alpha}$ with
$(0,h_{-}^{1,1})$. Then the only solution of $\kappa_{\alpha a b}
v^{\alpha} b^a b^b = 0$ that is meaningful (i.e. we cannot have
all $v^{\alpha} = 0$ as the Calabi-Yau manifold will vanish) is to
set $b^a = 0$ for all $a$. This is the only generic possibility
for a Minkowski vacuum in this case, depicted on Fig. \ref{hat}.

Note that $V = 0$ can be also achieved for $V_{CY} \rightarrow
\infty$, $e^{-\phi} = 0$, or $W = 0$. We are not interested in the
first two cases as these contradict our initial construction,
while $W = 0$ might be achieved for some solutions of $D_{z^i} W =
D_{\tau} W = 0$ (in this case $\kappa_{\alpha a b} v^{\alpha} b^a
b^b$ will be stabilized to a certain value since it appears in
$D_{\tau} W$). $W = 0$ will correspond to a supersymmetric
solution since then all covariant derivatives $D_I W$ vanish.
However, it is not clear how often this is possible since the
solutions of these equations cannot be given analytically, so the
only generic solution remains $b^a = 0$ for all $a$.

Therefore we are very restricted in terms of possible analytic
scenarios for stabilization of all moduli. The case when
$\kappa_{\alpha a b} v^{\alpha} b^a b^b = 0$ is a generic minimum
of the potential, corresponding to vanishing of all terms
dependent on the non-geometric K\"{a}hler moduli. This mechanism
leads us back to the no-scale potential that is flat in the
directions of the geometric K\"{a}hler moduli since $V = 0$ after
stabilizing all $b^a = 0$. Note that also the masses $m_{b^a} = 0$
in this case, which is not what we need as a final outcome.

So we need to improve our approach in order to break this no-scale
behavior and lift up the axion mass. At this point one can employ
the KKLT scenario of considering only D$3$-instantons and then
stabilizing all moduli at a supersymmetric point. A special case
of this idea was considered in \cite{stieb}. We will however stick
to the LVS procedure and calculate first the effect of the leading
perturbative corrections and only afterwards of the instanton
corrections on the potential that now includes the axionic moduli.
This is in fact the more general case and it does not exclude, but
only improves KKLT. Thus we will be able to consistently give mass
to the $b^a$'s and $c^a$'s in the general case without the need to
add D-terms in (\ref{full_potential}).

\section{\label{alpha'-corrections}Perturbative $\alpha'$-corrections}

Including the leading perturbative $\alpha'$-corrections as found
first\footnote{Strictly speaking, only the orientifold with
$h^{1,1}_- = 0$ was considered at first. Later it was shown in
\cite{grimm-louis} that this can be trivially extended for a generic
orientifold.} in \cite{BBHL} by reducing to $\mathcal{N} = 1$ the
results of \cite{bbhl-follow} for the $\mathcal{N} = 2$ case, the
K\"{a}hler potential becomes:
\begin{equation}\label{alpha'-corr kaehler once more}
 K = -\ln[i \int_{CY} \Omega (z) \wedge \bar{\Omega} (\bar{z})] -
  \ln (-i (\tau - \bar{\tau})) - 2 \ln\left( \frac{\kappa_{\alpha} v^{\alpha}}{6} + \frac{\xi}{2} \left(\frac{\tau-\bar{\tau}}{2 i} \right)^{3/2} \right),
\end{equation}
where $\xi$ is a constant, proportional to the Euler number of the
CY three-fold:
\begin{equation}\label{xi}
  \xi = - \frac{4 \chi \zeta (3)}{(2 \pi)^3}.
\end{equation}
There are no $\alpha'$-corrections to the superpotential in
perturbation theory and so $W$ is still given by (\ref{tree_W}).
Even only the addition of corrections in $K$ changes considerably
the potential as we will see shortly. $V$ does not have to be
positive semi-definite any more since $\xi$ could be either
positive or negative depending on the sign of the Euler number of
the Calabi-Yau. As we will see the sign of $\xi$ will directly
correspond to the sign of $V$.

To analyze the vacuum structure, we start again from
(\ref{full_potential}). The computation of the inverse K\"{a}hler
metric including the $\alpha'$-corrections is given in App.
\ref{appalpha} (see (\ref{inverse full kaehler metric alpha'
corrections}) and (\ref{prefactors})). Thus once more we obtain a
complicated expression\footnote{Nevertheless, Eq. (\ref{no-scale
potential - 4}) still holds. The factor $4$ is generic for this
class of K\"{a}hler potentials as discussed in \cite{grimm-PhD}.}
for the potential that cannot be minimized in a controlled way.
Similarly to the tree level case, we continue by imposing $D_{z^i}
W = D_{\tau} W = 0$. At tree level, the stabilization of the other
moduli then lead to minima at $V = 0$. This property does not hold
any more when the $\alpha'$-corrections are taken into account
since the potential is no longer bounded from below. In the
present case, we will know that we have found minima only if they
are at large volumes (in string units) $V_{CY}$ due to the
argument given in the Large Volume Scenario \cite{lvs1}. It goes
as follows. We write the full potential in a way to separate
clearly the contributions from $D_{z^i} W$ and $D_{\tau} W$ from
the other terms. So we split (\ref{full_potential}) in three terms
- a quadratic with respect to $D_{z^i} W, D_{\tau} W$ (both
summation indices in (\ref{full_potential}) run over $z^i, \tau$),
a linear (only one index including $z^i$ or $\tau$) and a constant
(both indices running over the other moduli). Further we focus on
the scaling of these terms with volume and thus we use the leading
terms of the inverse K\"{a}hler metric (\ref{inverse full kaehler
metric alpha' corrections to a satisfactory precision}):
\begin{eqnarray}\label{full_LVS_potential with axions}
\nonumber   V = e^K(K^{z^i \bar{z}^j} D_{z^i} W D_{\bar{z}^j}
\bar{W} + K^{\tau
\bar{\tau}} D_{\tau} W D_{\bar{\tau}} \bar{W})\\
+ \mathcal{O}(V_{CY}^{-2/3}) e^K (W D_{\bar{\tau}} \bar{W} +
\bar{W} D_{\tau} W) + V_{\alpha'},
\end{eqnarray}
with
\begin{equation}\label{alpha' corrected potential with v,b unstabilized}
  V_{\alpha'} = \frac{e^K e^{- 2 \phi} |W|^2}{4 V_{CY}} \left(3 \xi e^{\phi/2} + \frac{(\kappa_{\alpha a b} v^{\alpha} b^a b^b)^2}{V_{CY}} + \mathcal{O}(V_{CY}^{-2/3}) \right).
\end{equation}
In (\ref{alpha' corrected potential with v,b unstabilized}) we
have given only the leading terms in large volume, because the
complete analytic expression looks complicated
(c.f.(\ref{prefactors})) and we will only discuss large volume
stabilization for the following reason. The first term of
(\ref{full_LVS_potential with axions}) is positive semi-definite
and is only zero at the supersymmetric case $D_{z^i} W = D_{\tau}
W = 0$. This term dominates the other two at large volumes as it
scales as $V_{CY}^{-2}$ while the two others scale as
$V_{CY}^{-8/3}$ and $V_{CY}^{-3}$ respectively. Then any movement
of the complex structure and dilaton moduli away from the
supersymmetric point increases the potential, i.e. this point is a
stable minimum. The moduli potential simply becomes $V =
V_{\alpha'}$ and minimizing it with respect to $v^{\alpha}, b^a$
will result in full minimization of the initial moduli potential
in all directions as long as the large volume assumption is
satisfied for the obtained minima.

Therefore, we can consistently neglect the terms of order
$V_{CY}^{-11/3}$ and lower in (\ref{alpha' corrected potential
with v,b unstabilized}). We first observe that, apart from the
non-generic supersymmetric point at $W = 0$ (corresponding to KKLT
type of extremum), we again need to set $\kappa_{\alpha a b}
v^{\alpha} b^a b^b = 0 \Leftrightarrow \forall b^a = 0$ in order
to minimize the term depending on the axionic moduli. But in this
case we are still left with volume dependence since the $\xi$ term
survives. Now we see how important the sign of $\xi$ turns out to
be:

\begin{itemize}
  \item $\xi > 0$, i.e. $\chi_{CY} < 0$: The resulting potential is positive definite and vanishing as $V_{CY} \rightarrow
  \infty$, i.e. this case is consistent with our assumptions but leads to decompactification of the Calabi-Yau.
  One can only hope that non-perturbative effects will eventually create a minimum at some finite large value of
  the volume (this is what happens in the LVS).
  \item $\xi < 0$, i.e. $\chi_{CY} > 0$: In this case the minimum is when the volume goes to zero and the
  potential goes to $- \infty$. Clearly, none of these is in accordance with the approximations
  made so far, and we can only trust the result at large volumes
  where no minima can be found. Instanton corrections
  cannot help in generating large volume minima since they cannot uplift the global minimum at $V_{CY} = 0$.
  Therefore, this case is undesirable and one needs very different approach in order to solve the problem of
  stabilizing the moduli for positive Euler number Calabi-Yau three-folds.
\end{itemize}

\section{\label{D3-instantons}D-brane instanton corrections} Until now we only considered the tree-level
superpotential (\ref{tree_W}). Let us see what happens if we
assume that the compactification manifold meets the criteria that
allow for nonzero D$3$-instanton contributions to $W$.

At this point a few words about instantons are in order. In string
theory instantons can appear in Calabi-Yau compactifications when
Euclideanized branes wrap cycles of the manifold
\cite{beckerstrom}. If the branes wrap around cycles in such a way
that supersymmetry is preserved, the corresponding cycle is called
supersymmetric. It is exactly those cases that give a finite
non-vanishing contribution to some of the physical quantities. As
explained in \cite{witten}, the counting of zero modes for a
specific cycle eventually determines if it is supersymmetric or
not. This translates into a nontrivial condition on the given
cycle, depending on its dimension. For example (relevant here) it
turns out that the $4$-cycles that satisfy these criteria,
admitting D$3$-brane instantons, are the ones that have an Euler
number $\chi_E = 1$. However, this condition is more subtle after
the addition of fluxes \cite{renata,flux_instantons} and then one
has to check each cycle separately. Fundamental string worldsheets
as well as NS$5$-branes can also give rise to instantons. It turns
out that worldsheet instantons give rise to non-perturbative
$\alpha'$ corrections to the K\"{a}hler potential, while
D$3$-branes and NS$5$-branes contribute to the superpotential. In
this paper we shall neglect NS$5$ contributions since they are
subleading at large volume as discussed in \cite{hugo}.

The superpotential with D$3$-instanton corrections is then:
\begin{equation}\label{instanton corrected W once more}
  W = W_{tree} + \sum_{\alpha} A_{\alpha} (z^i, \tau, G^a) e^{-a_{\alpha} T_{\alpha}} = W_0 + W_{np},
\end{equation}
where the sum over $\alpha$ only goes through the supersymmetric
cycles. The coefficients $A_{\alpha}$ can in principle depend on
all other moduli except the $T_{\alpha}$'s but their explicit
dependence is hard to determine and does not lead to further
insight in the process of moduli stabilization at present (see,
e.g. section 2.4 of \cite{grimm}).

We can directly use the K\"{a}hler potential (\ref{alpha'-corr
kaehler once more}) since we already showed that the
$\alpha'$-corrections will substantially change the minimization
process and cannot be neglected. Therefore, the moduli potential
in analogy to (\ref{full_LVS_potential with axions}) will become:
\begin{eqnarray}\label{full_alpha' instanton_potential}
\nonumber & V = e^K(K^{z^{\alpha} \bar{z}^{\beta}} D_{z^{\alpha}}
W D_{\bar{z}^{\beta}} \bar{W} + K^{\tau
\bar{\tau}} D_{\tau} W D_{\bar{\tau}} \bar{W})\\
  & + \mathcal{O}(V_{CY}^{-2/3}) e^K (W D_{\bar{\tau}} \bar{W} + \bar{W} D_{\tau} W) + V_{\alpha'} + V_{np1} +
  V_{np2},
\end{eqnarray}
with
\begin{equation}\label{alpha' potential with v,b unstabilized}
  V_{\alpha'} = \frac{e^K e^{- 2 \phi} |W|^2}{4 V_{CY}} \left(3 \xi
e^{\phi/2} + \frac{(\kappa_{\alpha a b} v^{\alpha} b^a
b^b)^2}{V_{CY}} + \mathcal{O}(V_{CY}^{-2/3})
  \right),
\end{equation}
\begin{eqnarray}\label{np1 potential with v,b unstabilized}
\nonumber & V_{np1}  =  e^K \sum_{\alpha, \beta} \{ \left( -
\frac{3}{2} (\kappa \kappa_{\alpha \beta} - \frac{3}{2}
\kappa_{\alpha} \kappa_{\beta}) - \frac{3}{2} e^{- \phi} \kappa
\kappa^{a b} \kappa_{\alpha a c} b^c \kappa_{\beta b d} b^d +
\mathcal{O}(\kappa^0) \right) a_{\alpha} a_{\beta}
A_{\alpha} \bar{A}_{\beta} +\\
\nonumber & + (- i e^{- \phi} \kappa \kappa^{a b} \kappa_{\alpha b
c} b^c + \mathcal{O}(\kappa^0)) \left(a_{\alpha}
  A_{\alpha} \partial_{\bar{G}^a} \bar{A}_{\beta} -  a_{\alpha} \bar{A}_{\alpha} \partial_{G^a} A_{\beta} \right) + \\
 & + \left(- \frac{2}{3} e^{- \phi} \kappa
\kappa^{a b} + \mathcal{O}(\kappa^0) \right)
  \partial_{G^a} A_{\alpha} \partial_{\bar{G}^b} \bar{A}_{\beta} \}  e^{- (a_{\alpha} T_{\alpha} +
a_{\beta} \bar{T}_{\beta})},
\end{eqnarray}
\begin{equation}\label{np2 potential with v,b unstabilized}
 V_{np2} = e^K \sum_{\alpha} \left(\frac{3}{2} \kappa_{\alpha} + \mathcal{O}(\kappa^{-2/3})\right) \left(a_{\alpha} A_{\alpha} \bar{W} e^{-
a_{\alpha} T_{\alpha}} + a_{\alpha} \bar{A}_{\alpha} W e^{-
a_{\alpha} \bar{T}_{\alpha}} \right),
\end{equation}
where the summations are still only over supersymmetric cycles.
$V_{np1} = e^K K^{i \bar{j}} \partial_{i} W \partial_{\bar{j}}
\bar{W}$ and $V_{np2} = e^K K^{i \bar{j}} \left( K_i W
\partial_{\bar{j}} \bar{W} + \partial_{i} W K_{\bar{j}} \bar{W} \right), i,j = T_{\alpha}, G^a$ are new terms here - they
appear because now $\partial_{T_{\alpha}} W \neq 0$ and
$\partial_{G^a} W \neq 0$.

By the same argument from the discussion after Eq. (\ref{alpha'
corrected potential with v,b unstabilized}), at large volumes we
can consistently set $D_{z^{\alpha}} W = D_{\tau} W = 0$. The
resulting equations have $G^a$ and $T_{\alpha}$ dependence that is
suppressed with $V_{CY}$, so we can safely assume that all complex
structure moduli and the dilaton have been set to constants. Then,
\begin{equation}\label{final full alpha' instanton potential}
  V = V_{\alpha'} + V_{np1} + V_{np2}.
\end{equation}

In principle, at this point we can also choose to follow the KKLT
proposal and stabilize all moduli supersymmetrically, i.e.
requiring additionally $D_{T_{\alpha}} W = D_{G^a} W = 0$. The
solutions of these equations will correspond to a set of extrema
of the potential and one has to check explicitly which ones are
minima. Thus we would obtain a number of solutions to our problem
that unfortunately cannot be listed analytically and so we cannot
draw any further conclusions. Therefore we now turn to the LVS
idea of trying to minimize (\ref{final full alpha' instanton
potential}) at large manifold volume, which ensures us of finding
minima in the full moduli space.

However, it is not so straightforward to minimize (\ref{final full
alpha' instanton potential}) and we need to make some
simplifications of $V_{np1}$ and $V_{np2}$ in order to proceed.
The scaling of both expressions (\ref{np1 potential with v,b
unstabilized}) and (\ref{np2 potential with v,b unstabilized}) is
being dominated by the exponential terms, and more precisely by
the real part of the term in the exponent, while the imaginary
part decides on the sign. At large 4-cycle volumes the terms are
very suppressed and we can safely ignore them as the exponential
function drops to zero very rapidly\footnote{Strictly speaking, we
are cheating here. Even for the large cycles $\kappa_{L}$, big
enough values of $\kappa_{L a b} b^a b^b$ will make the
non-perturbative contributions important. We will neglect such
possibility at first and comment on it when we consider the
general case with many 4-cycles in subsection \ref{many smalls}.}.
Therefore the dominating terms in $V_{np1}$ and $V_{np2}$ will be
the ones corresponding to the small (supersymmetric) cycles
$\kappa_{\alpha}$, which we shall denote $\kappa_s$. Here we
implicitly assume that the internal manifold is of "Swiss-cheese"
type \cite{lvs2}, ensuring that small enough cycles do exist for
large overall volume. This is required so that the new terms
$V_{np1}$ and $V_{np2}$ can compete with the previously discussed
$V_{\alpha'}$ as otherwise D-instanton corrections are diminishing
and we arrive back at the situation of section
\ref{alpha'-corrections}.

\subsection{\label{1 small}One small 4-cycle}
Assuming for the moment that there is one small 4-cycle and all
the others are too big, in the sense that $e^{- \kappa_{\alpha}}
<< e^{- \kappa_s}$ for all $\alpha \neq s$, we finally obtain
\begin{equation}\label{roughly approximated potential}
V = e^K \left[ - \alpha (b) \kappa_s e^{-\frac{3}{4} a_s \kappa_s}
e^{\frac{3}{4} a_s e^{- \phi} \kappa_{s a b} b^a b^b} +
\frac{\beta (b)}{V_{CY}} + \gamma (b) (- \kappa_{s s}) V_{CY}
e^{-\frac{3}{2} a_s \kappa_s} e^{\frac{3}{2} a_s e^{- \phi}
\kappa_{s a b} b^a b^b} \right],
\end{equation}
where the exact dependence of $\alpha$, $\beta$ and $\gamma$ on
the $b^a$'s is coming from (\ref{alpha' potential with v,b
unstabilized}) - (\ref{np2 potential with v,b unstabilized}):
\begin{equation}\label{alpha of b}
\alpha (b) = - \frac{3}{2} \left(A_s \bar{W}  e^{- i a_s
(\frac{3}{2} \rho_s + \frac{3}{4} \kappa_{s a b} (C_o b^a - c^a)
b^b)} + \bar{A}_s W e^{i a_s (\frac{3}{2} \rho_s + \frac{3}{4}
\kappa_{s a b} (C_o b^a - c^a) b^b)} \right),
\end{equation}
\begin{equation}\label{beta of b}
  \beta (b) = \frac{e^{- 2 \phi} |W|^2}{4} \left(3 \xi e^{\phi/2} + \frac{(\kappa_{\alpha a b} v^{\alpha} b^a
  b^b)^2}{V_{CY}}
  \right),
\end{equation}
\begin{eqnarray}\label{gamma of b}
\nonumber & \gamma (b) = 6 [\left(\frac{3}{2} + \frac{3 e^{-
\phi}}{2 \kappa_{s s}} \kappa^{a b} \kappa_{s a c} b^c \kappa_{s b
d} b^d \right) a_s^2 |A_s|^2 + \frac{2 e^{- \phi}}{3 \kappa_{s s}}
\kappa^{a b}
\partial_{G^a} A_s
\partial_{\bar{G}^b} \bar{A}_s + \\
& + \frac{i e^{- \phi}}{\kappa_{s s}} \kappa^{a b} \kappa_{s b c}
b^c a_s \left(A_s
\partial_{\bar{G}^a} \bar{A}_s -
\bar{A}_s \partial_{G^a} A_s \right)].
\end{eqnarray}
We further need to assume $\kappa_{s s} \simeq - \sqrt{\kappa_s}$
a la LVS, in order to make sure the $\gamma$ term in (\ref{roughly
approximated potential}) is not subleading. This is the only
possibility to obtain large volume minima within the approximation
of neglecting multi-instanton contributions to (\ref{instanton
corrected W once more}), as proven in details in the Appendix of
\cite{proof}\footnote{Multi-instanton contributions can be safely
ignored as long as $a_s \kappa_s >> 1$ in string units.}.

In (\ref{alpha of b}) for the first time we explicitly see some
dependence on the moduli $\rho_s, c^a$ defined through
(\ref{non-geometric 4-form moduli}) and (\ref{non-geometric
moduli}). This means we are allowed to stabilize them in a way
that will maximize $\alpha$, thus minimizing the overall
potential. Since they appear only in the imaginary part of the
exponent they can only determine the sign of $\alpha$ but not its
magnitude (they can give a relative prefactor between $-1$ and
$1$). Therefore it is clear that $\rho_s, c^a$ arrange themselves
in a way to make the expression as large positive as possible.
Since they appear in the term $ a_s A_s \bar{W} e^{- i a_s
(\frac{3}{2} \rho_s + \frac{3}{4} \kappa_{s a b} (C_o b^a - c^a)
b^b)} + c.c.$, there will be one equation to constrain the
possible values of $\rho_s$ and the $c^a$'s. This will be enough
to stabilize $\rho_s$ as in the original LVS and the $c^a$'s still
remain unstabilized. Therefore $\alpha > 0$ with certainty and its
$b^a$-dependence is absorbed by $\rho_s$, such that
$$\alpha = 3 |A_s| |W|.$$

On the other hand, we know that $\gamma (b)$ must be positive as
it comes from the inner product of the vector $\partial_i W$ with
itself. $\beta (b)$ is also positive by assumption since a
negative value will not lead to consistent minima as shown in the
previous section. Then, in order to minimize the full potential,
the remaining free moduli will try to make the magnitude of the
terms with $\beta$ and $\gamma$ as small as possible and the
magnitude of the term with $\alpha$ as big as possible (as it
appears with negative sign).

To find the minima of the potential $V$ we need to solve the
system of equations $\frac{\partial V}{\partial b^a} = 0$ for all
$a$, $\frac{\partial V}{\partial \kappa_s} = 0$, and
$\frac{\partial V}{\partial V_{CY}} = 0$. To leading orders in
volume,
\begin{eqnarray}\label{minimizing even more roughly approximated potential}
\nonumber & \frac{\partial V}{\partial b^a} = e^K \kappa_{s a b}
b^b e^{-\frac{3}{4} a_s \kappa_s} e^{\frac{3}{4} a_s e^{- \phi}
\kappa_{s a b} b^a b^b} \sqrt{\kappa_s} \left[ - \frac{3}{2}
\alpha \sqrt{\kappa_s} + 3 \gamma V_{CY} e^{-\frac{3}{4} a_s
\kappa_s} e^{\frac{3}{4} a_s e^{- \phi} \kappa_{s a b} b^a b^b}
\right] + \\
&+ \frac{e^K e^{-2 \phi} |W|^2}{V_{CY}^2} \kappa_{\alpha a b}
v^{\alpha} b^b \kappa_{\alpha a b} v^{\alpha} b^a b^b + e^K
\frac{\partial \gamma}{\partial b^a} V_{CY} e^{-\frac{3}{2} a_s
\kappa_s} e^{\frac{3}{2} a_s e^{- \phi} \kappa_{s a b} b^a b^b}.
\end{eqnarray}
We have extrema of the potential in the $b$-moduli directions
whenever $\frac{\partial V}{\partial b^a} = 0$ for all $b^a$. This
is satisfied by $b^a = 0$ for all $a$\footnote{Here we further
assume that $\partial_{G^a} A_s = 0$ when all $b^a = 0$. Thus,
$\frac{\partial \gamma}{\partial b^a} = 0$ at this point of moduli
space. If this is not the case, $b^a = 0, \forall a$ cannot be an
extremum of $V$, but $\frac{\partial \gamma}{\partial b^a}$ will
still be small at large volumes and the extremum will be very
close to $b^a = 0, \forall a$ without changing our qualitative
discussion.}, while other solutions can be found only for specific
cases depending on the form of the intersection numbers
$\kappa_{\alpha a b}$ and the coefficients $A_s$.

Note that if all $b^a = 0$ we get back the Large Volume Scenario
\cite{lvs1}, $\alpha (b) = \alpha_{LVS}$, $\beta (b) =
\beta_{LVS}$, and $\gamma (b) = \gamma_{LVS}$. The solutions of
$\frac{\partial V}{\partial \kappa_s} = 0$ and $\frac{\partial
V}{\partial V_{CY}} = 0$ can be found explicitly only numerically,
but the small cycle will be always stabilized to $\kappa_s \approx
\ln (V_{CY})$. Then minima at large volume $V_{CY}$ can exist
under the same conditions as in \cite{lvs1,lvs2}, i.e. some
particular relative weight of the prefactors $\alpha$, $\beta$,
and $\gamma$ ($\beta
>> \alpha$ and/or $\gamma >> \alpha$)\footnote{It is easy to see from
(\ref{roughly approximated potential}) that the term with $\alpha$
will always dominate at $V_{CY} \rightarrow \infty$ as it will
scale additionally as $\ln (V_{CY})$, while the term with $\gamma$
only scales with $\sqrt{\ln (V_{CY})}$ and the term with $\beta$
has no additional scaling. Thus large volume minima can only be
reached when one of the positive $\beta$ and $\gamma$ terms
competes and dominates over the negative term until $V_{CY}$ is
large. So we can roughly estimate the relative weights of $\alpha,
\beta, \gamma$ based on scaling. We need $V_{CY} \gtrsim 10^6$,
thus $\beta \gtrsim 14 \alpha$ and/or $\gamma \gtrsim 4 \alpha$.
In the main text we denote these criteria $\beta
>> \alpha$ and/or $\gamma >> \alpha$ in order to keep the discussion as general as possible.}.

If all $b^a = 0$ we can go further and compute the matrix of
second derivatives:
\begin{equation}\label{sec derivatives roughly approximated potential}
\left( \frac{\partial^2 V}{\partial b^a \partial b^b} \right)_{b^a
= 0, \forall a}  = \frac{3 e^K}{V_{CY}} \left[ \kappa_{s a b}
\left( - \frac{1}{2} \alpha \ln (V_{CY}) + \gamma  \sqrt{\ln
(V_{CY})} \right) - \frac{e^{-\phi}}{2} \kappa^{c d} \kappa_{s a
c} \kappa_{s b d} \right].
\end{equation}
$\kappa^{c d} \kappa_{s a c} \kappa_{s b d}$ is a negative
definite matrix since $\kappa^{c d} = (\kappa_{c d})^{-1}$ (here
we use the definition $\kappa_{a b} \equiv \kappa_{\alpha a b}
v^{\alpha}$), so it always gives a positive contribution to
(\ref{sec derivatives roughly approximated potential}). However,
this term is subleading in $V_{CY}$ and therefore we concentrate
on the other part of the expression. In typical cases $\alpha$ and
$\gamma$ are of the same order of magnitude so the term in round
brackets is negative\footnote{$\alpha$ and $\gamma$ are determined
from the stabilization of $z^i$ and $\tau$, so "typical" here
refers to statistically more probable. $\gamma >> \alpha$ only
when $W_0$ is small.}. The matrix $\kappa_{s a b}$ could in some
cases be negative definite as it comes from the orientifold
projection and we know from before that $\kappa_{\alpha a b}
v^{\alpha}$ is negative definite. If this is the case, $\left(
\frac{\partial^2 V}{\partial b^a \partial b^b} \right)_{b^a = 0,
\forall a}$ is positive definite and therefore $b^a = 0$ is a
minimum of the potential with the Large Volume Scenario holding
for suitable values of $\alpha, \beta, \gamma$. In principle, even
when $\kappa_{s a b}$ has nonnegative eigenvalues we can have full
minima due to the positive contribution from $\kappa^{c d}
\kappa_{s a c} \kappa_{s b d}$ but this seems possible only for
not so large values of $V_{CY}$ and is therefore not a generic
case. The argument is reversed when $\gamma
>> \alpha$, as in this case the term in the round brackets becomes
positive and $b^a = 0$ is a minimum if $\kappa_{s a b}$ is
positive definite (not very likely).

Apart from this analytic class of minima, we can show the
existence of another class of minima, for which explicit solutions
cannot be given. From (\ref{minimizing even more roughly
approximated potential}) we see there can be extrema also when
$b^a \neq 0$ for some $a$'s. And, in fact, we know that some of
these extrema will certainly be minima of the potential as long as
$b^a = 0$ is not a minimum. The proof that there is always at
least one minimum of the potential is carried out in App.
\ref{proving}. It follows that if $\kappa_{s a b}$ does not
satisfy the above conditions to make $b^a = 0$ a minimum, then
there will still be a minimum with at least one of the $b^a$'s
nonzero. In this case we lose analytic control over the values of
$\kappa_s$ and $V_{CY}$ at the minimum, so we cannot a priori make
sure that the large volume assumption and the neglect of
multi-instanton contributions are justified. This will fully
depend on the explicit form of the intersection numbers
$\kappa_{\alpha a b}$. It is only clear that the $b^a$'s will
still tend to minimize $\kappa_{\alpha a b} v^{\alpha} b^a b^b$ in
(\ref{beta of b}), i.e. as many as possible of the $b$-moduli will
be zero if they are not fixed by $\frac{\partial V}{\partial b^a}
= 0$. One would naturally expect that the closer $\kappa_{s a b}
b^a b^b$ is to zero at the minimum, the closer the values of
$\kappa_s$ and $V_{CY}$ are to the LVS case.

To illustrate the above explicitly, consider a simple version of
(\ref{minimizing even more roughly approximated potential}) where
$\frac{\partial \beta}{\partial b^a}, \frac{\partial
\gamma}{\partial b^a}$ always vanish. Then another analytic
solution of $\frac{\partial V}{\partial b^a} = 0$ is $$\kappa_{s a
b} b^a b^b = \frac{4 e^{\phi}}{3 a_s} \ln \left( \frac{\alpha
\sqrt{\kappa_s}}{2 \gamma V_{CY}}\right) + e^{\phi} \kappa_s.$$
This is satisfied generally on a hypersurface of the full
$b$-moduli space where at least one of the $b^a$'s is nonzero.
Lower order corrections will then also fix the remaining free
$b^a$'s. It is easy to verify that this hypersurface is a minimum
in all $b$-directions, but when considering minimization in the
$\kappa_s$ and $V_{CY}$ directions this is no longer a valid
solution, as expected since our initial assumption to neglect the
$b$-dependence of $\beta$ and $\gamma$ is clearly wrong. However,
this gives us some intuition for what to expect roughly from the
possible minima that are not at $b^a = 0, \forall a$. It is likely
that brute-force solution of (\ref{minimizing even more roughly
approximated potential}) will only lead to a hypersurface of
minima that is subsequently refined by the lower order
corrections.

So finally we emerge with two main scenarios for stabilization of
the non-geometric moduli that entirely depend on the specific
Calabi-Yau intersection numbers. The two cases are sketchily
summarized in Table \ref{tab1}. If $\kappa_{s a b}$ has also zero
eigenvalues there will be flat directions at leading order, which
will be fixed by the subleading tree level term $\kappa_{\alpha a
b} v^{\alpha} b^a b^b$ in $\beta$ (we will see soon that for more
small moduli flat directions at leading order are unlikely to
appear).

\begin{table}
\begin{center}
\begin{tabular}{|c|c|c|c|} \hline
Generic minimum &  $\alpha \leftrightarrow \gamma$  &   Restr. on $\kappa_{s a b}$   & Large Volume  \\
    \hline
\multirow{2}{*}{$b^a = 0, \qquad \forall a$} & $\alpha \sqrt{\ln V_{CY}} > 2 \gamma$ & Neg. def. & $\beta >> \alpha$ \\
& $\alpha \sqrt{\ln V_{CY}} < 2 \gamma$ & Pos. def. & $\beta >> \alpha$ / $\gamma >> \alpha$ \\
\hline
$\frac{\partial V}{\partial b^a} = 0, \quad a = 1,..., h^{1,1}_-$ & $\alpha \sqrt{\ln V_{CY}} > 2 \gamma$ & Not neg. def. & $\beta >> \alpha$ \\
$\exists \tilde{a}$, s.t. $b^{\tilde{a}} \neq 0$ & $\alpha \sqrt{\ln V_{CY}} < 2 \gamma$ & Not pos. def. & $\beta >> \alpha$ / $\gamma >> \alpha$ \\
\hline
\end{tabular}
\caption{Axion stabilization scenarios for one small $4$-cycle
$s$.}\label{tab1}
\end{center}
\end{table}

\subsection{\label{many smalls}Many small 4-cycles}
Generalizing these conclusions for many small moduli is more
involved due to a subtlety coming from $V_{np1}$ (see (\ref{np1
potential with v,b unstabilized})). There we obtain a mix of
exponential terms for different cycles. Now also each separate
small four-cycle (as long as it is supersymmetric) will lead to a
corresponding non-perturbative contribution to the $\alpha$ and
$\gamma$ terms:
\begin{eqnarray}\label{roughly approximated potential many moduli}
\nonumber & V = e^K [ \frac{\beta (b)}{V_{CY}} + \sum_{i = 1}^n
\{- \alpha_i (b) \kappa_{s_i} e^{-\frac{3}{4} a_{s_i}
\kappa_{s_i}} e^{\frac{3}{4} a_{s_i} e^{- \phi} \kappa_{s_i a b}
b^a b^b} + \\
 & + \gamma_i (b) (-\kappa_{s_i s_i}) V_{CY}
e^{-\frac{3}{2} a_{s_i} \kappa_{s_i}} e^{\frac{3}{2} a_{s_i} e^{-
\phi} \kappa_{s_i
a b} b^a b^b}\} ] - \\
\nonumber & - \sum_{i < j} \{ 6 V_{CY} e^K [i e^{- \phi} \kappa^{a
b} \kappa_{s_i b c} b^c \left( a_{s_i}
  A_{s_i} \partial_{\bar{G}^a} \bar{A}_{s_j} -  a_{s_i} \bar{A}_{s_i} \partial_{G^a} A_{s_j} \right) + \frac{2}{3} e^{- \phi} \kappa^{a b}
  \partial_{G^a} A_{s_i} \partial_{\bar{G}^b} \bar{A}_{s_j} + \\
\nonumber & + \left(\frac{3}{2} \kappa_{s_i s_j} + \frac{3}{2}
e^{- \phi} \kappa^{a b} \kappa_{s_i a c} b^c \kappa_{s_j b d} b^d
\right) a_{s_i} a_{s_j} A_{s_i} \bar{A}_{s_j}] e^{- (a_{s_i}
T_{s_i} + a_{s_j} \bar{T}_{s_j})} + c.c. \},
\end{eqnarray}
where the constants $\alpha_i$ and $\gamma_i$ are defined in
analogy to (\ref{alpha of b}), (\ref{gamma of b}) with addition of
the index $i$ where needed to distinguish between different
small-volume cycles. Thus we can stabilize all $\rho_{s_i}$ by
maximizing each $\alpha_i$ separately. We see that the second part
of (\ref{roughly approximated potential many moduli}) (third and
fourth row) is a new term that mixes in a complicated way all
moduli $\kappa_{s_i}, \rho_{s_i}, b^a, c^a$ (hidden in the
exponents of $T_{s_i}$). Its value is ultimately restricted by the
condition $V_{np1} \geq 0$ so it must be smaller than the
$\gamma_i$ contributions. These additional terms will solve the
problem with the unstabilized $c^a$'s as they exhibit a nontrivial
dependence on them (unless all $b^a = 0$). Clearly, the minima
with respect to the $c^a$'s can only be found numerically after
specifying the concrete model and the number of stabilized axions
will depend on the values of $h^{1,1}_-$ and the cycles admitting
instanton corrections. Therefore the stabilization of $c^a$'s
cannot be controlled analytically very well, analogously to the
stabilization of $\rho_{\alpha}$'s.

As before, it is easy to see\footnote{Again, we assume that
$\partial_{G^a} A_{s_i} = 0$ for all $i$ when all $b^a = 0$.} that
there is an extremal point at $b^a = 0$ for all $a$: $\left(
\frac{\partial V}{\partial b^a} \right)_{b^a = 0, \forall a} = 0$.
Again, other analytic solutions of $\frac{\partial V}{\partial
b^a} = 0$ cannot be given, here the equation is even more
complicated than (\ref{minimizing even more roughly approximated
potential}). When all $b^a = 0$, we recover the many-cycle LVS.
The dependence on the $c^a$ moduli of the potential disappears
while dependence on $\rho_{s_i}$ becomes considerably more
complicated compared to the one small cycle case. A detailed
discussion of the minimization in the $\rho$-directions is given
in A.2 of \cite{proof} and we will not repeat it here. The main
result in the end is that large volume minima as before can still
exist for certain configurations of the intersection numbers (see
the reference for more details): again, $\kappa_{s_i} \approx \ln
(V_{CY})$, $(- \kappa_{s_i s_i}) \approx \sqrt{\ln (V_{CY})}$ for
all small cycles and the volume is stabilized at a large value.
This point is a minimum in the $b$-moduli directions as long as
\begin{equation}\label{minim V in b-directions with many small cycles}
\left( \frac{\partial^2 V}{\partial b^a \partial b^b} \right)_{b^a
= 0, \forall a} = \frac{3 e^K}{V_{CY}} \left[ \sum_{i=1}^n
\kappa_{s_i a b} \left( - \frac{1}{2} \alpha_i \ln (V_{CY}) +
\gamma_i \sqrt{\ln (V_{CY})} \right) + \mathcal{O} (V_{CY}^{-1/3})
\right]
\end{equation}
is positive definite. Note that in general $\alpha_1 \simeq
\alpha_2 ... \simeq \alpha_n$ and $\gamma_1 \simeq \gamma_2 ...
\simeq \gamma_n$ as they differ only by the small differences in
the proportionality constants $a_{s_i}, A_{s_i}$. Now the
condition for $b^a = 0$ to be minimum essentially states that the
combined matrix as sum of sub-matrices ($\kappa_{s_1 a b} + ... +
\kappa_{s_n a b}$)\footnote{Here we assumed exact equalities
$\alpha_1 = \alpha_2 ... = \alpha_n$ and $\gamma_1 = \gamma_2 ...
= \gamma_n$. Generally, the combined matrix of interest is a
weighted sum of $\kappa_{s_1 a b}, ..., \kappa_{s_n a b}$, but the
weights are nearly equal.} should be negative or positive definite
depending on the term in brackets (most likely $\alpha_i \sim
\gamma_i$ and then the matrix needs to be negative definite).

Once again, we can in general prove the existence of at least one
minimum. The argument goes exactly as in the case of one small
modulus in App. \ref{proving}, and the essential point will again
be that the dependence on $\gamma_i$ is not crucial asymptotically
so we can neglect it (this also means neglect of the additional
mixing terms as they arise together with the $\gamma_i$ terms from
(\ref{full_potential})). We will not repeat the same
considerations specifically for this case, as all statements in
App. \ref{proving} can be easily generalized to include many small
moduli.

There is now an important difference between the analytical
minimum $b^a = 0$ and the other possibility when at least one of
the $b^a$'s is nonzero. In the latter case the additional mixing
term will depend on the $c^a$'s and we will be able to fix some or
all of them while also being able to stabilize the $\rho_{s_i}$'s
in a more straightforward manner. The number of stabilized $c^a$'s
will depend on the number of nonzero $b$-fields so one needs to go
to the specific manifold model. Furthermore, when some $b^a$'s are
nonzero there is an additional subtlety. The full potential
originally depends also on the large cycle $\kappa_{L}$ but we
regarded this contribution as largely suppressed. However, if
$\kappa_{L a b} b^a b^b >> 0$ this assumption might not be correct
and there would be another term to consider. If this is the case
we can drop the requirement for "Swiss-cheese" manifold since we
will no longer make use of a clear distinction between small and
large 4-cycles. Again, this issue can only be assessed properly
once an explicit manifold is chosen.

Note that in order to obtain a large volume minimum, the $\gamma_i
>> \alpha_i$ option is questionable in general due to the fact
that the additional terms in (\ref{roughly approximated potential
many moduli}) could decrease substantially the effective value of
each $\gamma_i$. Therefore, in order to make sure $V_{CY}$ is
stabilized large in all cases, we require that $\beta >> \alpha_i,
\forall i$\footnote{As in the one small 4-cycle case, for the
non-analytic minima with nonzero $b^a$'s we cannot decide with
certainty about the criterion for large volume minimum. We can
only hope $\sum_{i} \kappa_{s_i a b} b^a b^b$ is not too far from
zero and then use the same requirement $\beta >> \alpha_i$ for all
$i$.}.

We have thus found possible minima not only for vanishing $b^a$'s,
but also for non-zero values. The main results are given in Table
\ref{tab2}. As we will see in the next section these minima could
be further destabilized by other instanton effects, so the
minimization of the axionic moduli turns out to be a nontrivial
step in the stabilization process.

\begin{table}
\begin{center}
\begin{tabular}{|c|c|c|c|} \hline
Generic minimum &  Restr. on $\sum_i \kappa_{s_i a b}$   & Large Volume & Stab. moduli  \\
    \hline
\multirow{2}{*}{$b^a = 0, \qquad \forall a$} &  \multirow{2}{*}{Neg. def.} & \multirow{2}{*}{$\beta >> \alpha_i, \forall i$} & \multirow{2}{*}{$z^i, \tau, V_{CY}, v^{s_i}, \rho_{s_i}$} \\
&  & & \\
\hline
$\frac{\partial V}{\partial b^a} = 0, \quad a = 1,..., h^{1,1}_-$ &  \multirow{2}{*}{Not neg. def.} & \multirow{2}{*}{$\beta >> \alpha_i, \forall i$} & $z^i, \tau, V_{CY}, v^{s_i}, \rho_{s_i},$ \\
$\exists \tilde{a}$, s.t. $b^{\tilde{a}} \neq 0$ &  &  & a part of $\{c^{a}\}$ \\
\hline
\end{tabular}
\caption{Axion stabilization scenarios for $n$ small $4$-cycles
$s_1...s_n$.}\label{tab2}
\end{center}
\end{table}

\section{\label{worldsheet instantons}Worldsheet instanton corrections} Another correction to the K\"{a}hler
potential is given by worldsheet instantons wrapping holomorphic
2-cycles on the internal manifold. It is inherited from the type
IIA $\mathcal{N} = 2$ prepotential \cite{mirror,bbhl-follow},
given by\footnote{Note the slight change of notation as compared
to \cite{mirror,bbhl-follow}. This is consistent with the intended
identification of the coordinates $X^I$ here and leads to the same
form of the K\"{a}hler potential.}:
\begin{equation}\label{prepotential}
    F_0 (X) = F_{cl} (X) + F_{pert} (X) + F_{non-pert} (X),
\end{equation}
with
\begin{eqnarray}\label{F_0}
\nonumber F_{cl} (X) = \frac{1}{3!} \frac{\kappa_{\alpha \beta
\gamma} X^{\alpha} X^{\beta} X^{\gamma}}{X^0}, \qquad F_{pert} = i
\frac{\xi}{8} (X^0)^2, \\
F_{non-pert} (X) = i \frac{(X^0)^2}{(2 \pi)^3}
\sum_{\Sigma_{\beta} \epsilon H_2} \sum_{n=1}^{\infty}
\frac{n_{\Sigma_{\beta}}^0}{n^3} e^{2 \pi n i k_{\alpha}^{\beta}
X^{\alpha}/X^0},
\end{eqnarray}
where $\xi$ is as defined in (\ref{xi}), $\kappa_{\alpha \beta
\gamma}$ are the usual intersection numbers, $k_{\alpha}^{\beta} =
\int_{\Sigma_{\beta}} \omega_{\hat{\alpha}}$, $2 X^{\alpha} \equiv
i v^{\hat{\alpha}} + b^{\hat{\alpha}}, \alpha = 1,...,h^{1,1}$
before orientifolding\footnote{The hats on $\hat{\alpha}$ are
introduced for a clear distinction between the $\mathcal{N} = 1$
variables of section \ref{flux} and the ones used here at
$\mathcal{N} = 2$.}, and $X^0$ has to be set to $1$ after
obtaining the K\"{a}hler potential. The numbers
$n_{\Sigma_{\beta}}^0$ are the genus zero topological invariants
of Gopakumar-Vafa \cite{gopakumar}, associated with each element
of the homology. Thus, $F_0$ is the prepotential for the vector
multiplets of type IIA at tree-level of string-loop expansion that
receives both perturbative and non-perturbative corrections in
$\alpha'$\footnote{These non-perturbative corrections are in fact
the genus zero worldsheet instanton contributions to the
prepotential. Higher genus worldsheets instantons do not appear in
the prepotential and will not be discussed further. Although it is
not fully precise, here we refer to the genus zero worldsheet
instantons simply as worldsheet instantons.}. It can be translated
to the type IIB orientifold picture by the classical c-map
\cite{c-map} with the new coordinates $X^{\alpha} = i
v^{\alpha}/2, \alpha = 1,...,h^{1,1}_+$ and $X^a = b^a/2, a =
1,...,h^{1,1}_-$ (see also e.g. \cite{uu} for more details on how this works). The relevant part of
the K\"{a}hler potential can then be calculated directly by
\begin{equation}\label{c-map keahler}
  K = - 2 \ln \left( i e^{-2 \phi} (X^I \frac{\partial \bar{F_0}}{\partial \bar{X^I}} - \bar{X^I} \frac{\partial F_0}{\partial
  X^I}) \right)_{X^0 = 1},
\end{equation}
where the final result for $K$ needs to be expressed as before in
terms of the chiral fields of section \ref{flux}. Using this, we
finally obtain in the Einstein frame
\begin{equation}\label{K_worldsheet instantons}
    K = -\ln(-i(\tau-\bar{\tau})) - 2 \ln \left( V_{CY} + \left(\frac{\tau-\bar{\tau}}{2 i} \right)^{3/2} \left( \frac{\xi}{2} + \frac{4}{(2 \pi)^3} \varpi_{ws} (\tau, G)
    \right) \right),
\end{equation}
with
\begin{equation}\label{F_term}
    \varpi_{ws} (\tau, G) = \sum_{\Sigma_{\beta} \epsilon H_2^-} \sum_{n=1}^{\infty} \frac{n_{\Sigma_{\beta}}^0}{n^3} \cos \left(n \pi \frac{k_a^{\beta} (G - \bar{G})^a}{\tau - \bar{\tau}} \right) = \sum_{\Sigma_{\beta} \epsilon H_2^-} \sum_{n=1}^{\infty} \frac{n_{\Sigma_{\beta}}^0}{n^3} \cos (n \pi k_a^{\beta} b^a),
\end{equation}
where $k_a^{\beta} = \int_{\Sigma_{\beta}} \omega_a$ and
$\omega_a$ are the harmonic $(1,1)$-forms that are odd under the
orientifold projection and $\Sigma_{\beta}$ are the corresponding
2-cycles. The contributions from the even $(1,1)$-forms are
exponentially suppressed with the $v^{\alpha}$'s and we can safely
neglect them. This new K\"{a}hler potential includes infinite
(converging) sum over $n$ and another sum over the elements of the
homology $H_2^-$ of the CY manifold\footnote{This sum also needs
to be finite, see section 2.2 of \cite{grimm} for discussion of
this issue.}. This makes the metric very hard to invert and we
cannot present a generic inverse of $K_{A \bar{B}}$ that is
manifold independent as was the case before. However, we can use
the intuition from previous results to draw quite generic
conclusion on how worldsheet instantons can influence the moduli
stabilization. Note that the corrections are subleading in volume,
\begin{equation}\label{K worldsheet approx}
    K = - 2 \ln (V_{CY} ) - 2  \frac{\left(\frac{\tau-\bar{\tau}}{2 i} \right)^{3/2} \left( \frac{\xi}{2} + \frac{4}{(2 \pi)^3} \varpi_{ws} (\tau, G)
    \right)}{V_{CY}} + O(V_{CY}^{-2}).
\end{equation}
Therefore the worldsheet instantons will appear in the end result
the same way as the perturbative corrections, i.e. in the
definition of the $\beta$ term (see Eqs.(\ref{roughly approximated
potential}) and (\ref{beta of b})). They would be too subleading
to influence the $\alpha, \gamma$ terms in (\ref{roughly
approximated potential}).

If we consider more closely the dependence of $\varpi_{ws}$, we
see that its extrema can be given by the condition $k_a^{\beta}
b^a = l^{\beta}$ for an integer number $l^{\beta}$. In fact
$\varpi_{ws}$ is maximized for $l^{\beta} = 0, \pm 2, \pm 4...$
and minimized whenever $l^{\beta} = \pm 1, \pm 3, \pm 5...$ for
every cycle $\Sigma_{\beta}$. If indeed $k_a^{\beta} b^a =
l^{\beta}$ for all $\Sigma_{\beta}$, then both $\frac{\partial
\varpi_{ws}}{\partial \tau}$ and $\frac{\partial
\varpi_{ws}}{\partial G^a}$ vanish. So in this case we can
effectively consider $\varpi_{ws}$ to be constant for the purpose
of obtaining analogs of (\ref{inverse full kaehler metric alpha'
corrections to a satisfactory precision}) and (\ref{break into
sums alpha corr}) that eventually determine the expression for
$\beta$. Then the K\"{a}hler metric can be again inverted
analytically\footnote{In the sense that the potential $V$ is
precise upto order $\mathcal{O} (V_{CY}^{-10/3})$, i.e. the
inverse metric is analytic at leading order. This is all we need
since we are working under the assumption of large volume. The
K\"{a}hler metric cannot be inverted to all orders due to the fact
that e.g. $\frac{\partial^2 \varpi_{ws}}{\partial G^a
\partial G^{\bar{b}}}$ does not vanish at the minimum.} just by
adding the constant $\frac{8}{(2 \pi)^3} \varpi_{ws}$ to the
existing $\xi$ in the formulae in App. \ref{appalpha}. For the
minimum of $\varpi_{ws}$ we get:
\begin{equation}\label{maximized F_term}
    \varpi_{ws} |_{k_a^{\beta} b^a = l^{\beta}, l^{\beta} = \pm 1, \pm 3..., \forall \Sigma_{\beta}} = \sum_{\Sigma_{\beta} \epsilon H_2^-} \sum_{n=1}^{\infty} \frac{(-1)^n n_{\Sigma_{\beta}}^0}{n^3} = -\frac{3 \zeta (3)}{4} \sum_{\Sigma_{\beta} \epsilon H_2^-}
    n_{\Sigma_{\beta}}^0.
\end{equation}
The minimization of $\varpi_{ws}$ means minimization of the full
potential as it decreases the value of the $\beta$ term:
\begin{equation}\label{beta corrected by worldsheet instantons}
  \beta_{ws} |_{k_a^{\beta} b^a = l^{\beta}, l^{\beta} = \pm 1, \pm 3..., \forall \Sigma_{\beta}} = \frac{3 e^K e^{- 3 \phi/2} |W|^2 \zeta (3)}{2 (2 \pi)^3} \left((-2 \chi) - 3 \sum_{\Sigma_{\beta} \epsilon H_2^-} n_{\Sigma_{\beta}}^0 \right).
\end{equation}
Note that in the case when $b^a = 0$ for all $a$, $\varpi_{ws}$ is
maximized and the sign in front of the instanton sum in (\ref{beta
corrected by worldsheet instantons}) flips. This will make $b^a =
0$ much less likely to be a minimum of the potential, e.g.
(\ref{sec derivatives roughly approximated potential}) and
(\ref{minim V in b-directions with many small cycles}) will be
corrected with the negative sum over Gopakumar-Vafa (GV)
invariants. In case it is large enough, the sum will make sure
that $b^a = 0, \forall a$ is in fact a maximum. And the minimum
will certainly be at a point which decreases $\beta$. If the
$\beta$ term decreases so much that it becomes negative we will no
longer have any consistent minima in the volume direction as
discussed in sections \ref{alpha'-corrections} and
\ref{D3-instantons}. $\beta
> 0$ is absolutely crucial for the existence of LVS minima, while
supersymmetric KKLT minima can exist independently of the sign of
$\beta$. On the other hand, if $\beta$ is positive of the order of
$\alpha, \gamma$ we will only have small volume minima in the LVS.
So one can only hope that the 2-cycles of the manifold do not
allow for larger values of the GV invariants and thus of the
worldsheet instanton corrections as this can spoil the whole
process of moduli stabilization. The case $\gamma >> \alpha$ might
still enable the existence of desired minima for small positive
$\beta$, but this does not seem to be possible for many
supersymmetric cycles as seen in subsection \ref{many smalls}.

The above discussion is in fact quite general and does not
necessarily have to hold only for the special points in $b$-moduli
space that minimize $\varpi_{ws}$, although these are the cases
that can be handled analytically (as long as the inverse
K\"{a}hler metric is concerned). Even for generic values of the
$b^a$'s at the minimum of the potential where we also get
corrections from $\frac{\partial \varpi_{ws}}{\partial G^a}$, the
$\beta$ term will tend to decrease as all terms in $\beta$ coming
from worldsheet corrections will necessarily be periodic and
therefore allowed to become negative\footnote{Unfortunately, as
long as $|2 \chi| < |3 \sum_{\Sigma_{\beta}}
n_{\Sigma_{\beta}}^0|$, $\beta$ will necessarily be stabilized
negative as this ensures the minimum of the potential is at very
large negative values $V_{min} \rightarrow - \infty$ and very
small manifold volumes. This is of course not an acceptable vacuum
as it contradicts all assumptions of our construction.}. Note that
in fact the worldsheet instantons are the leading term that
exhibits $b^a$-dependence. The tree-level term from section
\ref{tree_level} is suppressed by $V_{CY}^{1/3}$ compared to
$\varpi_{ws}$, and therefore we expect that (even without having
any D$3$-instantons) the $b^a$'s are stabilized away from zero,
unless the GV invariants are small. Thus the volume will be
usually stabilized at a lower value as compared to section
\ref{D3-instantons}, due to the decrease in the $\beta$ term. As
we see the risk for the stability of the LVS minima after adding
worldsheet instantons is very general and one needs to explicitly
calculate the invariants $n_{\Sigma_{\beta}}^0$ in order to make
sure phenomenologically accepted vacua are still present in a
given model. Even if this is so, the second type of generic vacuum
in Table \ref{tab2} is the most plausible (and least well
controlled) outcome.

\section{\label{CONC}Discussion}

We made some progress towards full stabilization of the scalar
fields in the compactification of type IIB string theory on
Calabi-Yau orientifolds with $h^{1,1}_- \neq 0$. As seen, the
search for supersymmetric and non-supersymmetric minima of the
moduli potential is a nontrivial task. Many approximations and
simplifications are employed in the process and it is not always
granted that these are justified in all possible models. Clearly,
perturbative and non-perturbative corrections play an important
role and it is unfortunate that at present there is no full
classification of possible terms that can appear in the K\"{a}hler
potential and the superpotential.

Nevertheless, in the literature one can find extensive discussion
of quantum corrections and their regime of importance, i.e. how
suppressed they are with the volume. References
\cite{jumpingloops}, \cite{BHK}, and \cite{proof} study this topic
in detail and show that string loop corrections for "Swiss-cheese"
CY manifolds are subleading compared to the perturbative and
non-perturbative $\alpha'$-corrections so they only help
stabilizing the non-supersymmetric 4-cycle volumes, but there may
be other types of manifolds for which this is not satisfied. Other
possible perturbative $\alpha'$-corrections are known to be less
important compared to the ones discussed in the LVS, i.e. it seems
that the LVS is safe from further perturbative $\alpha'$ and $g_s$
corrections. However, there might be other corrections from DBI
actions and $\mathcal{N}=1$ supergravity that are of importance
(see e.g. section 6 and 7 of \cite{jumpingloops} for more
details). If we are to claim that realistic string
compactifications have been found, a better understanding of all
quantum corrections is needed. Needless to say, same holds for
instanton corrections to $W$ and $K$ - as seen in the previous
section worldsheet instantons have the potential to break down the
LVS.

Despite these shortcomings, we managed to show with certainty that
all $b^a$'s are stabilized already at tree level with instanton
corrections possibly changing their vevs and lifting their masses,
and that the $\rho_{\alpha}$'s and (some of) the $c^a$'s are
stabilized if D$3$-instanton effects contribute to the moduli
potential of the given model. The above is true under the
condition that the manifold volume is stabilized large, which
ultimately depends on the topological data for the manifold and
the stabilization of the complex structure moduli that appear
implicitly in the $\alpha$, $\beta$, and $\gamma$ terms defined
through (\ref{roughly approximated potential})-(\ref{gamma of b}).
Therefore our procedure works for a subset of all minima that one
can find in the landscape of vacua, i.e. for those cases that
produce the desired relative weights of $\alpha, \beta, \gamma$ as
discussed in section \ref{D3-instantons}. How large this subset is
depends on the specific Calabi-Yau manifold, which also determines
the type of perturbative and non-perturbative corrections that
should be considered. So in the end everything can be determined
from the topological structure of the compactification manifold as
expected. It seems that at present the full generality of the
construction defined in section \ref{flux} is exhausted and one
needs to go to specific examples in order to obtain more explicit
results that can be used for predictive purposes.

We will now try to briefly describe some applications that make
use of these axionic moduli \cite{axion,conlon,axion-grimm}. In
type IIB the axions arise from the $2$-form fields $B_2$ and $C_2$
and the $4$-form field $C_4$ as given by Eqs. (\ref{non-geometric
moduli}) and (\ref{non-geometric 4-form moduli}). There are a few
ideas to employ these scalars for phenomenological purposes. One
scenario, developed initially by Peccei and Quinn
\cite{peccei-quinn}, proposes that a massive scalar field (an
axion) provides a solution to the CP problem in QCD. Reference
\cite{conlon} discusses in detail whether the missing Peccei-Quinn
axion can be coming from the $C_4$-moduli. The present work might
help answering this question also for the $B_2, C_2$-axions. To
study this, one should however also include open string moduli
which was beyond the scope of this paper. Another possibility to
use axions is for driving inflation in the early universe
\cite{N-flation} (also called N-flation). In \cite{axion-grimm}
and \cite{eva} the N-flation scenario with type IIB axions was
considered and made plausible in some specific toy models.
Therefore our work extends the possibility to study this idea as
it provides a more systematic approach to the subject. We leave
this for future research.

\section*{Acknowledgements} I am particularly grateful to my
supervisor Stefan Vandoren for his guidance during my master's
research and for many helpful discussions and comments on the
draft of this paper. I would also like to thank Frederik Denef and
Thomas Grimm for helpful correspondence. This work was supported
by the Huygens Scholarship Programme of the Netherlands
organization for international cooperation in higher education.

\appendix

\section{\label{AppB}Inverting the full K\"{a}hler metric}

Here we will present in detail how to deal with inverting the
matrix of partial derivatives $K_{I \bar{J}} =
\partial_I \partial_{\bar{J}} K$ after including the non-geometric moduli. We will not consider the complex structure
moduli dependence, as they are not coupled to the dilaton and the
other moduli in $K$:
\begin{equation}\label{metric split}
K_{I \bar{J}} = \left( \begin{array}{cc}
K_{z^i \bar{z}^j} & 0 \\
0 & K_{A \bar{B}}  \end{array} \right)
\end{equation}

$K_{z^i \bar{z}^j}$ cannot be inverted explicitly without a given
model, so we focus on inverting $K_{A \bar{B}}$. We will show
that, although rather non-trivial, there is an exact analytic
solution for the inverse metric $K^{A \bar{B}}$ both at tree level
and with perturbative $\alpha'$-corrections included. We will
therefore consider these cases separately in different
subsections.

\subsection{\label{tree appendix}Tree level} More explicitly, the relevant part of the K\"{a}hler potential (\ref{tree_kaehler}) is
\begin{equation}\label{tree kaehler yet once more}
 K = - \ln (-i (\tau - \bar{\tau})) - 2 \ln\left( \frac{(\frac{2}{3} (T_{\alpha} + \bar{T}_{\alpha}) - \frac{i}{2 (\tau - \bar{\tau})} \kappa_{\alpha a
b} (G-\bar{G})^a (G-\bar{G})^b) v^{\alpha} (T_{\alpha}, G^a,
\tau)}{6} \right),
\end{equation}
where $v^{\alpha}$ is an implicit function of the K\"{a}hler
coordinates. It is given by the relation $\kappa_{\alpha} =
\kappa_{\alpha \beta \gamma} v^{\beta} v^{\gamma} = \kappa_{\alpha
\beta} v^{\beta}$, where we made the definition $\kappa_{\alpha
\beta} \equiv \kappa_{\alpha \beta \gamma} v^{\gamma}$. Therefore,
$v^{\alpha} = \kappa^{\alpha \beta} \kappa_{\beta}$. We make an
analogous definition for
the intersection numbers with Latin indices: $\kappa_{a b} \equiv \kappa_{\alpha a b} v^{\alpha}$. \\
Now we can calculate the actual K\"{a}hler metric, using the
following matrix definitions that are used for shorthand and
easier calculation:
\begin{equation}\label{useful matrices}
G^{\alpha \beta} \equiv -\frac{2}{3} \kappa \kappa^{\alpha \beta}
+ 2 v^{\alpha} v^{\beta}, \qquad \qquad G_{a b} \equiv -
\frac{3}{2} \frac{\kappa_{a b}}{\kappa},
\end{equation}
and their corresponding inverses
\begin{equation}\label{useful matrix inverses}
G_{\alpha \beta} = -\frac{3}{2} \left( \frac{\kappa_{\alpha
\beta}}{\kappa} - \frac{3}{2} \frac{\kappa_{\alpha}
\kappa_{\beta}}{\kappa^2} \right), \qquad \qquad G^{a b} = -
\frac{2}{3} \kappa \kappa^{a b}.
\end{equation}
The first partial derivatives of $K$ can then be computed to be:
\begin{eqnarray}\label{partial_der_full_kaehler}
\nonumber K_{\tau} &=& - K_{\bar{\tau}} = - \frac{1}{\tau - \bar{\tau}} - \frac{3 i}{2 (\tau - \bar{\tau})^2 \kappa} \kappa_{a b} (G-\bar{G})^a (G-\bar{G})^b =\\
\nonumber &=& \frac{i e^{\phi}}{2} + i G_{a b} b^a b^b, \\
K_{G^a} &=& - K_{\bar{G}^a} = \frac{3 i}{(\tau - \bar{\tau}) \kappa} \kappa_{a b} (G-\bar{G})^b = 2 i G_{a b} b^b, \\
\nonumber K_{T_{\alpha}} &=& K_{\bar{T}_{\alpha}} = - \frac{2
v^{\alpha}}{\kappa},
\end{eqnarray}
where we used from (\ref{real Kaehler coordinates}) that $(\tau -
\bar{\tau}) = 2 i e^{- \phi}$ and $(G -
\bar{G})^a = - (\tau - \bar{\tau}) b^a = - 2 i e^{- \phi} b^a$.\\
Then,
\begin{eqnarray}\label{full keahler metric}
\nonumber K_{\tau \bar{\tau}} &=& \frac{e^{2 \phi}}{4} + e^{\phi} G_{a b} b^a b^b + \frac{9}{16 \kappa^2} G^{\alpha \beta} \kappa_{\alpha a b} b^a b^b \kappa_{\beta c d} b^c b^d,\\
\nonumber K_{G^a \bar{\tau}} = K_{\tau \bar{G}^a} &=& e^{\phi} G_{a b} b^b + \frac{9}{8 \kappa^2} G^{\alpha \beta} \kappa_{\alpha a b} b^b \kappa_{\beta c d} b^c b^d, \\
\nonumber K_{T_{\alpha} \bar{\tau}} = - K_{\tau \bar{T}_{\alpha}} &=& - \frac{3 i}{4 \kappa^2} G^{\alpha \beta} \kappa_{\beta a b} b^a b^b, \\
K_{G^a \bar{G}^b} &=& e^{\phi} G_{a b} + \frac{9}{4 \kappa^2} G^{\alpha \beta} \kappa_{\alpha a c} b^c \kappa_{\beta b d} b^d,            \\
\nonumber K_{T_{\alpha} \bar{G}^a} = - K_{G^a \bar{T}_{\alpha}} &=& - \frac{3 i}{2 \kappa^2} G^{\alpha \beta} \kappa_{\beta a b} b^b, \\
\nonumber K_{T_{\alpha} \bar{T}_{\beta}} &=& \frac{G^{\alpha
\beta}}{\kappa^2}.
\end{eqnarray}
The inverse metric can be found after a lengthy calculation, which
goes as follows. One can make an ansatz for each of the elements
of the inverse metric from the number of free indices, e.g. the
component $K^{T_{\alpha} \bar{\tau}}$ has only one free lower
index $\alpha$ as opposed to the upper index of the original
metric component. Therefore, a possible ansatz could be
$K^{T_{\alpha} \bar{\tau}} = a \kappa_{\alpha} + b \kappa_{\alpha
a b} b^a b^b$, where $a$ and $b$ can be any expression with fully
contracted indices (or with no indices at all). Plugging the
ansatz for every element of the inverse matrix leads to 9 coupled
equations which lead to unique determination of all components.
For the specific example, we find $a = 0$, $b = 3 i e^{- \phi}$.
Explicitly, the whole inverse metric is:
\begin{eqnarray}\label{inverse full kaehler metric}
\nonumber K^{\tau \bar{\tau}} &=& 4 e^{- 2 \phi}, \\
\nonumber K^{G^a \bar{\tau}} = K^{\tau \bar{G}^a} &=& - 4 e^{- 2 \phi} b^a, \\
\nonumber K^{T_{\alpha} \bar{\tau}} = - K^{\tau \bar{T}_{\alpha}} &=& 3 i e^{- 2 \phi} \kappa_{\alpha a b} b^a b^b, \\
K^{G^a \bar{G}^b} &=& e^{- \phi} G^{a b} + 4 e^{- 2 \phi} b^a b^b,  \\
\nonumber K^{T_{\alpha} \bar{G}^a} = - K^{G^a \bar{T}_{\alpha}} &=& -\frac{3 i e^{- \phi}}{2} G^{a b} \kappa_{\alpha b c} b^c - 3 i e^{- 2 \phi} \kappa_{\alpha b c} b^b b^c b^a,  \\
\nonumber K^{T_{\alpha} \bar{T}_{\beta}} &=& \kappa^2 G_{\alpha
\beta} + \frac{9 e^{- \phi}}{4} G^{a b} \kappa_{\alpha a c} b^c
\kappa_{\beta b d} b^d + \frac{9 e^{-2 \phi}}{4} \kappa_{\alpha a
b} b^a b^b \kappa_{\beta c d} b^c b^d.
\end{eqnarray}
Having found the inverse metric and the first partial derivatives
(\ref{partial_der_full_kaehler}), to obtain Eq. (\ref{no-scale
potential - 4}) is down to some trivial algebra. However, it might
be quite interesting in which way one gets the number $4$. We can
break up the sum into two parts - a sum which runs over all
$T_{\alpha}$ and $G^a$ but not over the dilaton, plus the
remainder of the whole sum (i.e. where at least one of the indices
goes over $\tau$). Then,
\begin{equation}\label{break into sums}
K^{i \bar{j}} K_i K_{\bar{j}} = 3 + 9 e^{- 2 \phi} \left(
\frac{\kappa_{\alpha a b} v^{\alpha} b^a b^b}{\kappa} \right)^2,
\qquad \qquad i,j = T_1...T_{h_{+}^{(1,1)}},
G^1...G^{h_{-}^{(1,1)}},
\end{equation}
while for the remainder one gets $1 - 9 e^{- 2 \phi} \left(
\frac{\kappa_{\alpha a b} v^{\alpha} b^a b^b}{\kappa} \right)^2$
as expected since the two sums add up to $4$.

\subsection{Perturbative $\alpha'$-corrections\label{appalpha}} In order to simplify notation, we first use the following definitions:
\begin{eqnarray}\label{define Y}
\nonumber \hat{\xi} & \equiv & \frac{\xi}{2 (2 i)^{3/2}}, \\
Y & \equiv & V_{CY} + \frac{\xi}{2} \left(\frac{\tau-\bar{\tau}}{2
i} \right)^{3/2} = \frac{\kappa}{6} + \hat{\xi} (\tau -
\bar{\tau})^{3/2}.
\end{eqnarray}
In the following we will drop the hat of $\hat{\xi}$ and will use
this new definition until the end of the section where we switch
to the proper definition. With these identifications, the
K\"{a}hler potential takes a misleadingly simple form:
\begin{equation}\label{full kaehler potential with alpa' corrections}
K = - \ln (-i (\tau - \bar{\tau})) - \ln (Y).
\end{equation}
However, $Y$ is now dependent on all variables. Its partial
derivatives are:
\begin{eqnarray}
\nonumber \frac{\partial Y}{\partial T_{\alpha}} = \frac{\partial Y}{\partial \bar{T}_{\alpha}} &=& \frac{v^{\alpha}}{6}\\
\frac{\partial Y}{\partial G^a} = - \frac{\partial Y}{\partial \bar{G}^a} &=& - \frac{i}{4 (\tau - \bar{\tau})} \kappa_{a b} (G - \bar{G})^b  \\
\nonumber \frac{\partial Y}{\partial \tau} = - \frac{\partial
Y}{\partial \bar{\tau}} &=& \frac{i}{8 (\tau - \bar{\tau})^2}
\kappa_{a b} (G - \bar{G})^a (G - \bar{G})^b + \frac{3}{2} \xi
(\tau - \bar{\tau})^{1/2},
\end{eqnarray}
where we define $\kappa_{a b}, \kappa_{\alpha \beta}$ as in the
previous subsection. However, we slightly change the definition of
$G^{\alpha \beta}, G_{a b}$:
\begin{equation}\label{useful matrices alpha correction}
G^{\alpha \beta} \equiv -\frac{Y}{9} \kappa^{\alpha \beta} +
\frac{1}{18} v^{\alpha} v^{\beta}, \qquad \qquad G_{a b} \equiv -
\frac{\kappa_{a b}}{4 Y}.
\end{equation}
Their corresponding inverses are
\begin{equation}\label{useful matrix inverses alpha correction}
G_{\alpha \beta} = - \frac{9 \kappa_{\alpha \beta}}{Y} +
\frac{\kappa_{\alpha} \kappa_{\beta}}{2 Y \left( - \frac{Y}{9} +
\frac{\kappa}{18} \right)}, \qquad \qquad G^{a b} = - 4 Y
\kappa^{a b}.
\end{equation}
With these, and using $(\tau - \bar{\tau}) = 2 i e^{- \phi}$, $(G
- \bar{G})^a = - (\tau - \bar{\tau}) b^a = - 2 i e^{- \phi} b^a$,
\begin{eqnarray}\label{partial_der_full_kaehler alpha' corrections}
\nonumber K_{\tau} = - K_{\bar{\tau}} &=& - \frac{1}{\tau - \bar{\tau}} - \frac{i}{4 (\tau - \bar{\tau})^2 Y} \kappa_{a b} (G-\bar{G})^a (G-\bar{G})^b -\\
\nonumber - \frac{3 \xi (\tau - \bar{\tau})^{1/2}}{Y} &=& \frac{i e^{\phi}}{2} + i G_{a b} b^a b^b - \frac{3 \xi (2 i)^{1/2} e^{- \phi/2}}{Y}, \\
K_{G^a} = - K_{\bar{G}^a} &=& \frac{i}{2 (\tau - \bar{\tau}) Y} \kappa_{a b} (G-\bar{G})^b = 2 i G_{a b} b^b, \\
\nonumber K_{T_{\alpha}} = K_{\bar{T}_{\alpha}} &=& -
\frac{v^{\alpha}}{3 Y}.
\end{eqnarray}
The metric then takes the form:
\begin{eqnarray}\label{full keahler metric alpha' corrections}
\nonumber K_{\tau \bar{\tau}} &=& \left( \frac{e^{2 \phi}}{4} + \frac{3 \xi e^{\phi/2}}{2 (2 i)^{1/2} Y} - \frac{9 i \xi^2 e^{- \phi}}{Y^2} \right) +\\
\nonumber &+& \left( e^{\phi} + \frac{3 i \xi (2 i)^{1/2} e^{- \phi/2}}{Y} \right) G_{a b} b^a b^b + \frac{9}{16 Y^2} G^{\alpha \beta} \kappa_{\alpha a b} b^a b^b \kappa_{\beta c d} b^c b^d,\\
\nonumber K_{G^a \bar{\tau}} = K_{\tau \bar{G}^a} &=& \left( e^{\phi} + \frac{3 i \xi (2 i)^{1/2} e^{- \phi/2}}{Y} \right) G_{a b} b^b +\\
\nonumber &+& \frac{9}{8 Y^2} G^{\alpha \beta} \kappa_{\alpha a b} b^b \kappa_{\beta c d} b^c b^d, \\
\nonumber K_{T_{\alpha} \bar{\tau}} = - K_{\tau \bar{T}_{\alpha}} &=& - \frac{3 i}{4 Y^2} G^{\alpha \beta} \kappa_{\beta a b} b^a b^b - \frac{\xi (2 i)^{1/2} e^{- \phi/2}}{2 Y^2} v^{\alpha}, \\
K_{G^a \bar{G}^b} &=& e^{\phi} G_{a b} + \frac{9}{4 Y^2} G^{\alpha \beta} \kappa_{\alpha a c} b^c \kappa_{\beta b d} b^d,            \\
\nonumber K_{T_{\alpha} \bar{G}^a} = - K_{G^a \bar{T}_{\alpha}} &=& - \frac{3 i}{2 Y^2} G^{\alpha \beta} \kappa_{\beta a b} b^b, \\
\nonumber K_{T_{\alpha} \bar{T}_{\beta}} &=& \frac{G^{\alpha
\beta}}{Y^2}.
\end{eqnarray}
The inverse metric is found along the procedure from the previous
subsection, described after Eq.(\ref{full keahler metric}). For
easier reading, we will write down the ansatz for the inverse
metric and give the resulting prefactors separately.
\begin{eqnarray}\label{inverse full kaehler metric alpha' corrections}
\nonumber K^{\tau \bar{\tau}} &=& a, \\
\nonumber K^{G^a \bar{\tau}} = K^{\tau \bar{G}^a} &=& b b^a, \\
\nonumber K^{T_{\alpha} \bar{\tau}} = - K^{\tau \bar{T}_{\alpha}} &=& c \kappa_{\alpha a b} b^a b^b + d \kappa_{\alpha}, \\
K^{G^a \bar{G}^b} &=& l G^{a b} + m b^a b^b,  \\
\nonumber K^{T_{\alpha} \bar{G}^a} = - K^{G^a \bar{T}_{\alpha}} &=& e G^{a b} \kappa_{\alpha b c} b^c + f \kappa_{\alpha b c} b^b b^c b^a + q \kappa_{\alpha} b^a,  \\
\nonumber K^{T_{\alpha} \bar{T}_{\beta}} &=& g G_{\alpha \beta} +
h G^{a b} \kappa_{\alpha a c} b^c \kappa_{\beta
b d} b^d + j_1 \kappa_{\alpha a b} b^a b^b \kappa_{\beta c d} b^c b^d +\\
\nonumber &+& j_2 (\kappa_{\alpha} \kappa_{\beta a b} b^a b^b +
\kappa_{\alpha a b} b^a b^b \kappa_{\beta}) + j_4 \kappa_{\alpha}
\kappa_{\beta}.
\end{eqnarray}
The corresponding prefactors are
\begin{eqnarray}\label{prefactors}
\nonumber a &=& \frac{2 \left( - \frac{Y}{9} + \frac{\kappa}{18} \right)}{2 \left( - \frac{Y}{9} + \frac{\kappa}{18} \right) \left( \frac{e^{2 \phi}}{4} + \frac{3 \xi e^{\phi/2}}{2 (2 i)^{1/2} Y} - \frac{9 i \xi^2 e^{- \phi}}{Y^2} \right) + \frac{i \kappa \xi^2 e^{- \phi}}{Y^2}} \\
\nonumber b &=& - a \\
\nonumber c &=& \frac{3 i \left( - \frac{Y}{9} + \frac{\kappa}{18} \right)}{2 \left( 2 \left( - \frac{Y}{9} + \frac{\kappa}{18} \right) \left( \frac{e^{2 \phi}}{4} + \frac{3 \xi e^{\phi/2}}{2 (2 i)^{1/2} Y} - \frac{9 i \xi^2 e^{- \phi}}{Y^2} \right) + \frac{i \kappa \xi^2 e^{- \phi}}{Y^2} \right)} \\
\nonumber \\
\nonumber d &=& - \frac{(2 i)^{1/2} \xi e^{- \phi/2}}{2 \left( - \frac{Y}{9} + \frac{\kappa}{18} \right) \left( \frac{e^{2 \phi}}{4} + \frac{3 \xi e^{\phi/2}}{2 (2 i)^{1/2} Y} - \frac{9 i \xi^2 e^{- \phi}}{Y^2} \right) + \frac{i \kappa \xi^2 e^{- \phi}}{Y^2}} \\
\nonumber e &=& - \frac{3 i}{2} e^{- \phi} \\
\nonumber f &=& - c \\
\nonumber q &=& - d \\
g &=& Y^2 \\
\nonumber h &=& \frac{9}{4} e^{- \phi} \\
\nonumber j_1 &=& \frac{9 \left( - \frac{Y}{9} + \frac{\kappa}{18}
\right)}{8 \left( 2 \left( - \frac{Y}{9} + \frac{\kappa}{18}
\right) \left( \frac{e^{2 \phi}}{4} + \frac{3 \xi e^{\phi/2}}{2 (2
i)^{1/2} Y} - \frac{9 i
\xi^2 e^{- \phi}}{Y^2} \right) + \frac{i \kappa \xi^2 e^{- \phi}}{Y^2} \right)} \\
\nonumber \\
\nonumber j_2 &=& \frac{3 i \xi (2 i)^{1/2} e^{- \phi/2}}{4 \left(
2 \left( - \frac{Y}{9} + \frac{\kappa}{18} \right) \left(
\frac{e^{2 \phi}}{4} + \frac{3 \xi e^{\phi/2}}{2 (2 i)^{1/2} Y} -
\frac{9 i \xi^2 e^{-
\phi}}{Y^2} \right) + \frac{i \kappa \xi^2 e^{- \phi}}{Y^2} \right)} \\
\nonumber \\
\nonumber j_4 &=& - \frac{2 i \xi^2 e^{- \phi} }{\left( -
\frac{Y}{9} + \frac{\kappa}{18} \right) \left( 2 \left( -
\frac{Y}{9} + \frac{\kappa}{18} \right) \left( \frac{e^{2
\phi}}{4} + \frac{3 \xi e^{\phi/2}}{2 (2 i)^{1/2} Y}
- \frac{9 i \xi^2 e^{- \phi}}{Y^2} \right) + \frac{i \kappa \xi^2 e^{- \phi}}{Y^2} \right)} \\
\nonumber l &=& e^{- \phi} \\
\nonumber m &=& a.
\end{eqnarray}
Clearly the inverse metric in this form is not very suitable for
calculational purposes. As we are interested in the large volume
behavior, we can expand the coefficients $a,...,m$ and take the
leading terms in the limit where $V_{CY} \rightarrow \infty$. In
order to calculate $K^{i \bar{j}} K_i K_{\bar{j}}$ exactly upto
$\mathcal{O}(V_{CY}^{-5/3})$ we also need some of the subleading
terms of the inverse metric. With this choice of relevant
accuracy, we obtain:
\begin{eqnarray}\label{inverse full kaehler metric alpha' corrections to a satisfactory precision}
\nonumber K^{\tau \bar{\tau}} & \approx & 4 e^{- 2 \phi} - \frac{24 \xi e^{-7 \phi/2}}{(2 i)^{1/2} V_{CY}}, \\
\nonumber K^{G^a \bar{\tau}} = K^{\tau \bar{G}^a} & \approx & - 4 e^{- 2 \phi} b^a, \\
\nonumber K^{T_{\alpha} \bar{\tau}} = - K^{\tau \bar{T}_{\alpha}} & \approx & 3 i e^{- 2 \phi} \kappa_{\alpha a b} b^a b^b - \frac{9 \xi (2 i)^{1/2} e^{- 5 \phi/2}}{V_{CY}} \kappa_{\alpha}, \\
K^{G^a \bar{G}^b} & \approx & e^{- \phi} G^{a b} + 4 e^{- 2 \phi} b^a b^b,  \\
\nonumber K^{T_{\alpha} \bar{G}^a} = - K^{G^a \bar{T}_{\alpha}} & \approx & -\frac{3 i e^{- \phi}}{2} G^{a b} \kappa_{\alpha b c} b^c -\\
\nonumber &-& 3 i e^{- 2 \phi} \kappa_{\alpha b c} b^b b^c b^a + \frac{9 \xi (2 i)^{1/2} e^{- 5 \phi/2}}{V_{CY}} \kappa_{\alpha} b^a,  \\
\nonumber K^{T_{\alpha} \bar{T}_{\beta}} & \approx & Y^2 G_{\alpha
\beta} + \frac{9 e^{- \phi}}{4} G^{a b} \kappa_{\alpha a c} b^c
\kappa_{\beta
b d} b^d + \frac{9 e^{- 2 \phi}}{4} \kappa_{\alpha a b} b^a b^b \kappa_{\beta c d} b^c b^d +\\
\nonumber &+& \frac{27 i \xi (2 i)^{1/2} e^{-5 \phi/2}}{4 V_{CY}}
(\kappa_{\alpha} \kappa_{\beta a b} b^a b^b + \kappa_{\alpha a b}
b^a b^b \kappa_{\beta}) - \frac{81 i \xi^2 e^{- 3 \phi}}{V_{CY}^2}
\kappa_{\alpha} \kappa_{\beta}.
\end{eqnarray}
Now we can calculate $K^{i \bar{j}} K_i K_{\bar{j}}$ and we find
it again equal to $4$ as in the tree level case (\ref{no-scale
potential - 4}). This time the $4$ comes as follows:
\begin{eqnarray}\label{break into sums alpha corr}
\nonumber K^{i \bar{j}} K_i K_{\bar{j}} &=& 3 + \frac{e^{- 2 \phi}
|W|^2}{4 V_{CY}^2} (\kappa_{\alpha a b} v^{\alpha}
b^a b^b)^2 - \frac{6 \hat{\xi} e^{-3 \phi/2}}{(2 i)^{1/2} V_{CY}} + \mathcal{O}(V_{CY}^{-5/3}),\\
i,j &=& T_1...T_{h_{+}^{(1,1)}}, G^1...G^{h_{-}^{(1,1)}},
\end{eqnarray}
and the remainder is what is left such that the sum is $4$. Note
that we still have $\hat{\xi}$ dependence, and if we switch to
$\xi$ we recover the standard term that appears in the literature
(c.f. (17) of \cite{lvs1}):
$$- \frac{6 \hat{\xi} e^{-3 \phi/2}}{(2 i)^{1/2} V_{CY}} = \frac{3 \xi e^{-3 \phi/2}}{4 V_{CY}}.$$

\section{\label{proving}Proof for the existence of minima of the moduli potential in the $b^a$-directions}
Here we present an extensive argument to show that the moduli
potential $V$ in its form (\ref{roughly approximated potential})
will always exhibit at least one minimum with respect to the
moduli in question, $\kappa_s, V_{CY},$ and $h^{1,1}_-$ $b^a$'s.
The argument can be trivially extended for the case of many small
4-cycles $\kappa_{s_i}$ (c.f. Eq. (\ref{roughly approximated
potential many moduli})).

Apart from the coefficients $\alpha, \beta, \gamma$, we see that
the potential $V$ (\ref{roughly approximated potential}) depends
on $b^a$ only through $\kappa_{s a b} b^a b^b$. Therefore, let us
define $x \equiv \kappa_{s a b} b^a b^b$ and take $V$ as a
function of only $x, \kappa_s, V_{CY}$. We will then consider all
possible cases of scaling of $\beta$ and $\gamma$ with $x$
($\alpha$ does not really depend on $x$ since any change in the
$b^a$'s only changes the value of $\rho_s$ at its minimum and
leaves $\alpha$ the same). So,

\begin{equation}\label{V of x}
    V (x, \kappa_s, V_{CY}) = - \alpha \frac{\kappa_s e^{- \kappa_s}
    e^{x}}{V_{CY}^2} + \frac{\beta (x)}{V_{CY}^3} + \frac{\gamma (x) \sqrt{\kappa_s} e^{-2 \kappa_s} e^{2
    x}}{V_{CY}},
\end{equation}
where we implicitly absorbed additional constants in the
definitions of $\alpha, \beta, \gamma, \kappa_s, x$ in order to
simplify notation and without any loss of generality. Note that
here $\alpha, \beta, \gamma$ are always strictly positive, while
$x\in (-\infty, +\infty)$ (for a completely generic matrix
$\kappa_{s a b}$) and $\kappa_s, V_{CY}\in (0, +\infty)$. Since
$V:\mathbb{R} \times \mathbb{R}^+ \times \mathbb{R}^+ \mapsto
\mathbb{R}$ we cannot really picture it, but we will present three
slices of $V$ at different values of $x$: $0$ and $\pm \infty$.
Then we will be able to draw conclusions on how the potential
looks everywhere.

\begin{itemize}
  \item $x \rightarrow - \infty$: \\
    $$ \lim_{x \rightarrow - \infty} V = \frac{\beta (x \rightarrow
    -\infty)}{V_{CY}^3}. $$
    Since $\beta (x \rightarrow -\infty)$ is always positive (in this limit it is in fact going to positive infinity), we see that for all values of $\kappa_s, V_{CY}$
    the potential remains positive and vanishes from above when $V_{CY} \rightarrow \infty$.

  \item $x = 0$: \\
    $$V = - \alpha_{LVS} \frac{\kappa_s e^{- \kappa_s}}{V_{CY}^2} + \frac{\beta_{LVS}}{V_{CY}^3} +
    \frac{\gamma_{LVS} \sqrt{\kappa_s} e^{-2 \kappa_s}}{V_{CY}}.$$
    Here the standard LVS is reproduced, the potential
    $V$ is large positive for small values of $\kappa_s, V_{CY}$,
    then goes below zero as they increase, and approaches zero asymptotically
    from below as $V_{CY} \rightarrow \infty$. The minimum of the
    potential is at $\kappa_s \approx \ln (V_{CY})$ and some
    finite value of $V_{CY}$ that depends on the coefficients and
    is not relevant for the argument here.

  \item $x \rightarrow \infty$: \\
    $$ \lim_{x \rightarrow \infty} V = - \alpha \frac{\kappa_s e^{- \kappa_s} e^{\infty}}{V_{CY}^2} + \frac{\beta (x \rightarrow
    \infty)}{V_{CY}^3}+\frac{\gamma (x \rightarrow
    \infty)  \sqrt{\kappa_s} e^{-2 \kappa_s} e^{2 \infty}}{V_{CY}}.$$
    This case is particularly subtle and we need to split it into
    a few subcases.

    For finite $\kappa_s$ we see that the last term is largely
    dominant as it rises squarely faster than the first term. This
    makes the potential always positive for finite values of the
    volume. When $V_{CY} \rightarrow 0$ the second term will
    make sure the potential never goes negative.

    On the other hand, when $\kappa_s$ goes faster to $\infty$
    than $x$ the second term will dominate everywhere and the potential is positive and only vanishing as $V_{CY} \rightarrow \infty$.

    The
    most subtle case is when $\kappa_s$ goes to infinity together with $x$. Then, $e^{x - \kappa_s}$ will remain finite
    and
    $$\lim_{ \kappa_s \rightarrow x, x \rightarrow  \infty} V = \lim_{x \rightarrow  \infty}  - \alpha \frac{x}{V_{CY}^2} +
    \frac{\beta (x)}{V_{CY}^3}+\frac{\gamma (x)\sqrt{x}}{V_{CY}}.$$
    Consider $\beta = const. + \frac{(\kappa_{a b} b^a
    b^b)^2}{V_{CY}}$. $x \rightarrow \infty$ only when at least one of the $b^a$'s goes to infinity.
    But then, since $\kappa_{a b}$ is negative definite, $(\kappa_{a b} b^a
    b^b)^2$ will necessarily also become infinite. Therefore, in the limit $x \rightarrow \infty$ the second term
    will
    always scale as $\frac{x^2}{V_{CY}^{10/3}}$ (remember that $\kappa_{a b} \equiv \kappa_{\alpha a b} v^{\alpha}$).
    The scaling of $\gamma (x)$ is less clear, but this is not important for our
    argument. We can even neglect the $\gamma$ term completely (since in any case it gives a positive contribution) and
    still prove our point. The potential is then simply $-\frac{\alpha'}{V_{CY}^2} +
    \frac{\beta'}{V_{CY}^{10/3}}$ with the $x$-dependence hidden in $\alpha' \sim x,
    \beta' \sim x^2$. Then it is straightforward to minimize the potential
    in the volume direction, and the value of the potential
    at the minimum is $$V_{min} = -\frac{6 \sqrt{3} \alpha'^{5/2}}{25 \sqrt{5} \beta'^{3/2}}
    \sim
    -\frac{1}{\sqrt{x}} \Rightarrow \lim_{x \rightarrow \infty} V_{min} = 0.$$ The minimum of the potential increases
    with $x$, so although $V$ can be (infinitesimally) negative, its real
    minimum will not be at $x \rightarrow \infty$ but at some
    finite value of $x$.
\end{itemize}

We have exhausted the limiting cases and showed there is no
runaway direction for $x$ and it must remain finite in order to
minimize $V$. And for finite fixed $x = x_0$ the potential will
have no runaway directions in the $\kappa_s$ and $V_{CY}$
directions. This is the case because the standard LVS ($x_0 = 0$)
behavior of the potential will still hold, only that for $x_0 \neq
0$ the relative weight between the coefficients $\alpha_{eff}
\equiv \alpha e^{x_0}, \beta_{eff} = \beta (x_0), \gamma_{eff} =
\gamma (x_0) e^{2 x_0}$ will change, effectively changing the
values of $\kappa_s, V_{CY}$ at the corresponding AdS minimum.
Since $\beta_{eff}$ always remains positive, $\kappa_s, V_{CY}$
will be finite at the minimum and the minimum itself will be at a
finite value $V_{min}$ (c.f. Appendix A of \cite{proof} for a
detailed proof).

Now we can safely claim that in all cases the full potential $V(x,
\kappa_s, V_{CY})$ will be minimized at a point or points inside
the domain of the variables, i.e. $x, \kappa_s, V_{CY}$ will all
have finite values at the minima. We are unable to specify the
number of minima, but we know there will be at least one since the
minimum of the potential is finite and negative at the LVS slice
$x = 0, \kappa_s \approx \ln (V_{CY})$, while on the boundaries of
its domain it is positive or vanishing. This concludes our proof
for the existence of minima of (\ref{roughly approximated
potential}).

\end{document}